%% file: main.tex
\documentclass[journal, final]{IEEEtran}
\IEEEoverridecommandlockouts

\usepackage[ruled,vlined]{algorithm2e}

\usepackage{amsmath,amssymb,capt-of}

\usepackage[utf8x]{inputenc}
\usepackage[super]{nth}

\usepackage{cite}

\usepackage[acronym,toc]{glossaries}

\newcommand{\ie}{i.e.,~}
\newcommand{\eg}{e.g.,~}

\usepackage[inline]{enumitem}

\usepackage[table]{xcolor}

\usepackage{array}

\usepackage{amsmath}

\usepackage{caption}
\usepackage{subcaption}
\usepackage{float}

\usepackage{lastpage,fancyhdr,graphicx}

\usepackage{multirow}
\usepackage{tabularx}

\usepackage{etoolbox}
\makeatletter
\patchcmd{\@verbatim}
  {\verbatim@font}
  {\verbatim@font\small}
  {}{}
\makeatother

\usepackage{makecell}

\usepackage{pifont}
\definecolor{ao}{rgb}{0.0, 0.5, 0.0}
\definecolor{amber}{rgb}{1.0, 0.49, 0.0}
\newcommand{\yestick}{{\color{ao}\ding{51}}}
\newcommand{\notick}{{\color{red}\ding{55}}}

\usepackage{nameref,hyperref}

\newcommand{\bots}{Likely Bots}
\newcommand{\semibots}{Likely Semi-Bots}
\newcommand{\humans}{Likely Humans}

\begin{document}

\title{BOTTER: A framework to analyze\\ social bots in Twitter}

%\author{
%    \IEEEauthorblockN{Javier Pastor-Galindo\IEEEauthorrefmark{1}, F\'elix G\'omez M\'armol\IEEEauthorrefmark{1}}, Gregorio Mart\'inez P\'erez\IEEEauthorrefmark{1}\\
%    \IEEEauthorblockA{\IEEEauthorrefmark{1}\textit{Department of Information and Communications Engineering, University of Murcia}, 30100 Murcia, Spain
%    \\\{javierpg, felixgm, gregorio\}@um.es}\\
%}

\author{
    \IEEEauthorblockN{Javier Pastor-Galindo, F\'elix G\'omez M\'armol, Gregorio Mart\'inez P\'erez}\\
    \IEEEauthorblockA{\textit{Department of Information and Communications Engineering, University of Murcia}, 30100 Murcia, Spain
    \\\{javierpg, felixgm, gregorio\}@um.es}\\
}

\maketitle

\begin{abstract}
Social networks have triumphed in communicating with people online, but they have also been exploited to launch influence operations for manipulating society. For that purpose, the deployment of software-controlled accounts (\eg social bots) has proven to be one of the most effective enablers. While powerful tools have been developed and adopted for their detection, the way to characterize these accounts and measure their impact is heterogeneous in the literature. To unify the existing efforts, we propose BOTTER, a common framework to analyze the interference of social bots in Twitter. The methodology compares the non-authentic actors with the rest of the users from diverse perspectives, thus building objective metrics to highlight differences. We validate the framework by applying it to a dataset of Twitter iterations dated in the weeks preceding the 2019 Spanish general election. In this sense, we check that our framework facilitates the quantitative evaluation of unauthentic groups, particularly discovering that social bots changed the natural dynamics of the network these days but did not have a significant impact. It also constitutes a practical tool for the qualitative interpretation of experimental results, particularly suggesting that semi-automated accounts are potentially more threatening than fully automated ones within the aforementioned electoral context.
\end{abstract}

\begin{IEEEkeywords}
Data analysis, Open data, Social networking (online), Bot (Internet), Internet security
\end{IEEEkeywords}

\input{introduction.tex}
\input{sota.tex}
\input{framework/0-preliminaries}

\input{framework/1-groups}
\input{framework/2-distribution}

\input{framework/3-composition}
\input{framework/4-robustness}
\input{framework/5-spreading}
\input{framework/6-slicing}

\input{framework/7-timeline}

\input{framework/8-virality}
\input{experiments.tex}
\input{discussion}

\input{conclusions.tex}

\section*{Acknowledgments}
This study was partially funded by the Spanish Government grants FPU18/00304 and RYC-2015-18210, co-funded by the European Social Fund. Authors would also like to acknowledge Prof. Karl Aberer at EPFL, Prof. Albert Blarer at Armasuisse, H\'ector Cordob\'es and IMDEA Networks Institute for their support to this work.

\bibliographystyle{IEEEtran}
\bibliography{biblio}
\end{document}

%% file: introduction.tex
\section{Introduction}\label{introduction}

Social media has become an ideal arena for communicating with a large audience and freely transmitting ideas, thoughts, and news~\cite{pastor2020}. While this has brought numerous benefits to society, it has also given rise to malicious operations that steer public opinion and manipulate users by spreading misinformation and deploying fake accounts~\cite{9451574}. 

Some of these illicit online activities specifically have electoral contexts as their primary target~\cite{9226407}, which are ideal gaps where to maximize the influence political behaviour and decision-making~\cite{Stephan2020}. The danger this poses for democracy and freedom of thought has mobilized major companies (e.g., Facebook, Instagram, Twitter, YouTube, and TikTok) to fight against influence operations\footnote{about.fb.com/news/2021/05/influence-operations-threat-report}, platform manipulation\footnote{transparency.twitter.com/en/reports/platform-manipulation}, integrity breakdowns\footnote{newsroom.tiktok.com/en-us/tiktoks-h-2-2020-transparency-report}, or misinformation spread\footnote{support.google.com/youtube/answer/9891785?hl=en}, but have not yet achieved a mature defence strategy~\cite{NATOStratComCOE2020}.

Diverse cutting-edge research lines are nowadays focusing on complementing these industry efforts. We could highlight strong lines in measuring of social media manipulation~\cite{Aral858}, analysis of content to counter influence operations~\cite{Alizadeh2020}, analysis of networks to infer misleading information~\cite{Pierri2020}, or detection of propaganda~\cite{Martino2020}. 

Additionally, academia and industry are obtaining significant advances in the fight against social bots~\cite{10.1145/3409116}, those semi or fully automated profiles that programmatically launch interactions on Online Social Networks (OSN)~\cite{8808170}. Although there are chat, informative, or joke social bots~\cite{electronics9111779}, most studies tend to spot malicious fake accounts that create numerous iterations and alter the natural dynamics of social media~\cite{Caldarelli2020}. In this sense, academic evidence mainly focuses on detecting these illegitimate accounts, performing high-level statistical and temporal analysis to compare with legitimate users~\cite{10.1007/978-3-030-05414-4_34}. However, they lack clear methodologies to infer the real impact of such social bots~\cite{10.1145/3409116} and encompass the influence of their actions~\cite{starbird2019disinformation}.

The paper at hand presents BOTTER, a data-oriented framework to analyze the impact of social bots in Twitter by comparing them with legitimate users from multiple perspectives. The differences between illegitimate and legitimate accounts on measured metrics facilitate the quantification and evaluation of the malicious interference. To validate the proposed framework, we apply it to assess the impact that semi and fully-automated bots had in the context of the 2019 Spanish general election.

The remainder of the paper is organized as follows. Section~\ref{sota} compiles several recent and remarkable studies that include analysis of social bots in social networks. %Section~\ref{framework-design} introduces the proposed framework by extracting its most significant properties.
Section~\ref{sec:framework-phases} describes the framework and formally defines their perspectives. Section~\ref{experiments} presents the experimental results of applying the methodology to the 2019 Spanish electoral context, which are accordingly interpreted and discussed in Section~\ref{discussion}. Finally, Section~\ref{conclusions} includes highlighted remarks derived from this research work and suggests some future working lines.

%% file: sota.tex
\section{Efforts to analyze social bots in the literature}
\label{sota}

\begin{table*}[h!]
\newcolumntype{C}[1]{>{\centering\let\newline\\\arraybackslash\hspace{0pt}}m{#1}}
%\color{blue}
\begin{tabular}{|| C{2.4cm} || C{1.6cm} | C{1.2cm} | C{1.1cm} | C{1.4cm} | C{1.2cm} | C{1.3cm} | C{1.3cm} | C{1cm} | C{1.3cm} ||} 
 \hline
 \textbf{Work} & \textbf{Comparative} & \textbf{Statistical} & \textbf{Network} & \textbf{Robustness} & \textbf{Influence} & \textbf{Structure} & \textbf{Temporal} & \textbf{Content} & \textbf{Virality} \\
\hline \hline

2016 Stweeler framework~\cite{10.1145/2872518.2889360} & \yestick & \yestick & \notick & \notick & \notick & \notick & \notick  & \yestick  & \notick  \\ \hline

2016-2017 Articles monitoring~\cite{Shao2017} & \yestick & \yestick & \yestick & \yestick & \yestick & \notick & \notick  & \yestick  & \yestick \\ \hline

2016 Brexit Botnet~\cite{Bastos2017} &  \yestick & \yestick & \yestick & \notick & \notick & \notick & \yestick  & \yestick  & \yestick  \\ \hline
 
2016 US Election~\cite{10.1007/978-3-319-58559-8_30} & \yestick & \yestick & \notick & \notick & \notick & \notick & \yestick  & \yestick  & \notick  \\ \hline
 
2014-2017 US Vaccine debate~\cite{10.2105/AJPH.2018.304567} & \yestick & \yestick & \notick & \notick & \notick & \notick & \notick  & \yestick  & \notick  \\ \hline

2016 Large scale bots analysis~\cite{Gilani2018} & \yestick & \yestick & \notick & \yestick & \notick & \notick & \notick  & \yestick  & \notick  \\ \hline

2018 Swedish Election~\cite{8587347} & \yestick & \yestick & \yestick & \notick & \notick & \notick & \notick  & \yestick  & \notick  \\ \hline
 
2016 Major Events~\cite{Schuchard2019} & \yestick & \yestick & \yestick & \notick & \yestick & \notick & \yestick  & \notick  & \notick  \\ \hline
 
2016 Russian Interference~\cite{Badawy2019} & \yestick & \notick & \yestick & \notick & \notick & \yestick & \yestick  & \yestick  & \notick  \\ \hline

2016 Three global events~\cite{10.1007/978-3-030-05414-4_34} & \yestick & \yestick & \yestick & \notick & \yestick & \notick & \notick  & \notick  & \notick  \\ \hline

2018 US Election~\cite{Luceri2019} & \yestick & \yestick & \yestick & \notick & \notick & \yestick & \notick  & \yestick  & \notick  \\ \hline

2016-2017 Russian Trolls~\cite{10.1145/3308560.3316495} & \yestick & \yestick & \notick & \notick & \notick & \notick & \yestick  & \yestick  & \notick  \\ \hline

2017-2018 Four riots events~\cite{Kusen2020} & \yestick & \yestick & \yestick & \notick & \notick & \notick & \yestick  & \yestick  & \notick  \\ \hline
 
2019 ES Election~\cite{9226407} & \yestick & \yestick & \notick & \notick & \notick & \notick & \yestick  & \yestick  & \notick  \\ \hline 
 
2018 US Election~\cite{Id2021} & \yestick & \yestick & \yestick & \notick & \yestick & \notick & \notick  & \notick  & \notick  \\ \hline \hline

\textbf{Our framework} & \yestick & \yestick & \yestick & \yestick & \yestick & \yestick & \yestick  & \yestick  & \yestick  \\

\hline
\end{tabular}
\caption{Analysis perspectives of social bots in related works}
\label{table:sota}
\end{table*}

Research on social bots boosted about a decade ago~\cite{10.1145/3409116}, mainly focusing on the great challenge of identifying these fake actors among the large volume of existing accounts~\cite{Yang_Varol_Hui_Menczer_2020}. The community has built several tools and models for bot detection ever since, widely employed by numerous studies to evidence the manipulation of online platforms and characterize the identified illegitimate groups. 

The impact of fake accounts has been measured in different online ecosystems. In a study of the \textit{StackExchange} portal~\cite{10.1145/3041021.3051116}, the authors applied a network-based framework to evaluate the activity dynamics of trolls, inferring that small groups of trolls specially affected the participation of legitimate users of the periphery, and large groups of trolls drastically reduced the activity of users in the core network. Another investigation on \textit{Twitch}~\cite{10.1145/3018661.3018672} counted the messages from both chatbots and users, demonstrating that chatbots sent far more traffic volume. In another research launched on the music platform \textit{SoundCloud}~\cite{quteprints121660}, the comments, friendships, and network stats of social bots were inspected to conclude that these fake accounts aggressively posted repetitive comments and had a small number of followers. Social bots have also been evaluated in \textit{Facebook} through the response time, comment length, or the number of posted links, determining that their activity rate was fifty times higher than expected~\cite{Santia_Mujib_Williams_2019}.

As seen from the above, each use case focuses on analyzing bots and comparing to legitimate users from particular perspectives. In the context of Twitter social bots, we have compiled in Table~\ref{table:sota} a group of studies that evaluate their behaviour, impact, or interference, categorizing them according to the type of analysis applied. As a common denominator, we find that the compiled works follow a \textit{comparative perspective} to contextualize social bots regarding the non-bots ones (frequently considered as humans)~\cite{10.1145/2872518.2889360, Shao2017, Bastos2017, 10.1007/978-3-319-58559-8_30, 10.2105/AJPH.2018.304567, Gilani2018, 8587347, Schuchard2019, Badawy2019, 10.1007/978-3-030-05414-4_34, Luceri2019, 10.1145/3308560.3316495, Kusen2020, 9226407, Id2021}. Generally, these studies perform \textit{statistical analysis} of the number of existing accounts, the traffic generated, or the interactions between groups. We also find that \textit{content analysis} is commonly applied to characterize the narratives of legitimate and illegitimate actors, identifying hashtags, extracting topics, parsing URLs, or inferring sentiments and emotions~\cite{10.1145/2872518.2889360, Shao2017, Bastos2017, 10.1007/978-3-319-58559-8_30, 10.2105/AJPH.2018.304567, Gilani2018, 8587347, Badawy2019, Luceri2019, 10.1145/3308560.3316495, Kusen2020, 9226407}. 

Secondly, there is a tendency in some projects to translate the case studies into graphs to apply \textit{network analysis}, leading to the construction of the retweet network and calculation of metrics such as the number of nodes, edges, strengths, betweenness, or closeness~\cite{Shao2017, Bastos2017, 8587347, Schuchard2019, Badawy2019, 10.1007/978-3-030-05414-4_34, Luceri2019, Kusen2020, Id2021}. The \textit{temporal analysis} is also characteristic in half of the selected studies, in which actions and behavior of users are observed over time to obtain temporal patterns or expose anomalies~\cite{Bastos2017, 10.1007/978-3-319-58559-8_30, Schuchard2019, Badawy2019, 10.1145/3308560.3316495, Kusen2020, 9226407}.

The rest of the proposed perspectives are not that frequent in the literature but no less important either. Two of the analyzed researches carry out a \textit{robustness analysis} of the social graph by disconnecting bots and measuring the impact of their existence in the network~\cite{Shao2017,Gilani2018}. On the other hand, four works address the complex challenge of \textit{influence analysis} by extracting centrality measures or applying well-known algorithms such as PageRank or HITS~\cite{Shao2017, Schuchard2019, 10.1007/978-3-030-05414-4_34, Id2021}. Moreover, it is less common to find \textit{structure analysis} applied to the social graph or retweet network by decomposing them in layers to reveal properties of both the periphery or core network~\cite{Badawy2019, Luceri2019}. Finally, only two works make a \textit{virality analysis} of tweets to in-depth inspect the diffusion cascades and understand content propagation over time~\cite{Shao2017, Bastos2017}.

As observed in the comparison table, the current works are individually unique and combine different perspectives. Despite the unifying comparison conducted, each project uses its own nomenclature and designs its specific experiments, evidencing an unclear and heterogeneous methodology for analyzing the phenomena around fake accounts. The latter does not mean that the research is wrongly conducted, but social bot findings obtained from different research methodologies are hard to compare and extrapolate to build \textit{social bot science} jointly.

%It is worth highlighting that a team proposed an agent-based model to analyze the influence of manipulative actors in social networks~\cite{doi:10.1080/0960085X.2018.1560920}, but consists on a theoretical work built on a political communication theory that is not aligned with the literature efforts of analyzing social bots. 

The paper at hand addresses this challenge by designing a clearly defined methodology to measure the impact of social bots, assembling the most common perspectives adopted in the literature. To the best of our knowledge, such a proposed framework is the first effort to harmonize the post-identification analysis.

%% file: framework/0-preliminaries.tex
\section{Proposed framework to analyze social bots in Twitter}\label{sec:framework-phases}

BOTTER defines a guided methodology to measure the interference of certain users (\eg social bots, fake news spreaders, or a specific subset of accounts) on social network dynamics in relation with the rest of the actors. This comparative perspective exposes malicious patterns and corrupted properties that suggest whether or not the targeted groups have impacted and influenced the system. %The framework has been designed to fulfil the following properties

\subsection*{Preliminaries}

The proposed framework demands a snapshot of a set of Twitter interactions between users within a specific time range from $d_{1}$ to $d_{m}$. An interaction \textit{t} is composed of the following metadata: retweeted user ($t^{retweeted}$), the user who retweets ($t^{retweeter}$), the tweet that is retweeted ($t^{tweet}$), the topics of the retweeted tweet ($t^{topics}$), and the timestamp of the retweet ($t^{timestamp}$).

\begin{figure}[ht]
    \centering
    \includegraphics[width=\columnwidth]{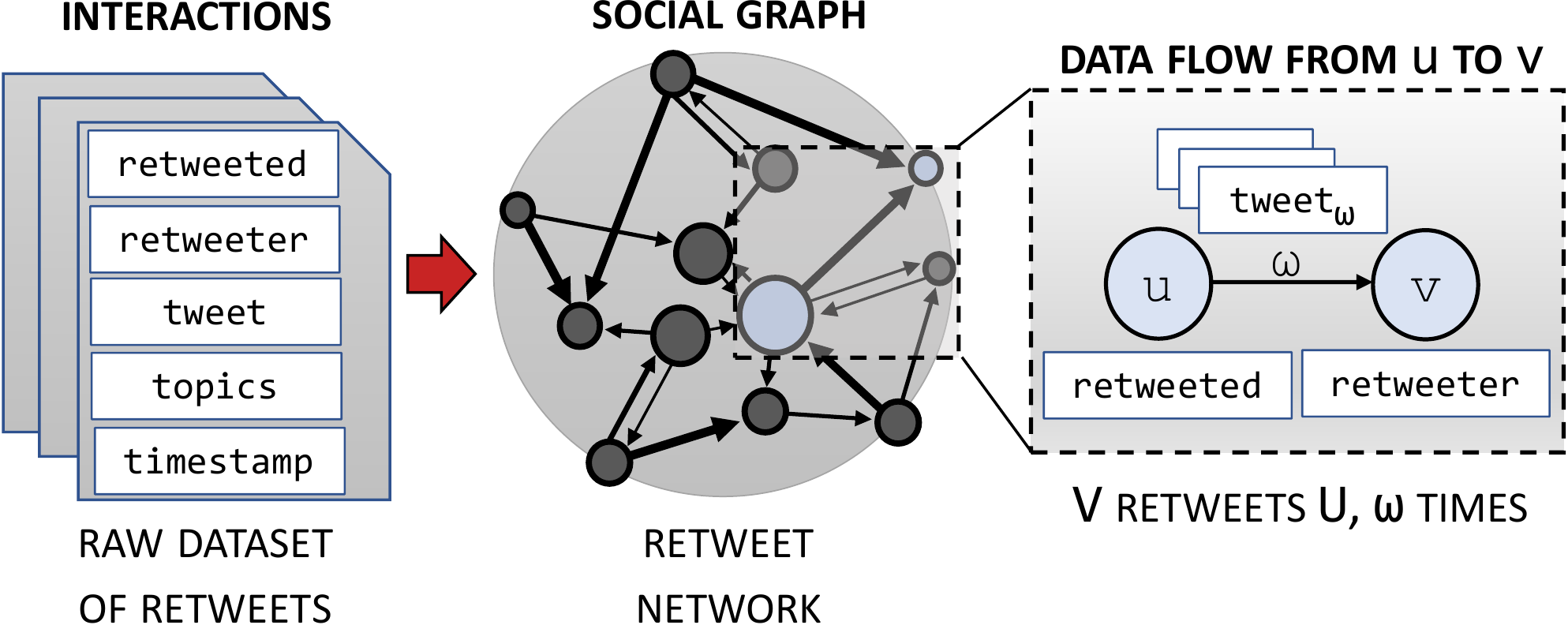}
    \caption{Social graph from interactions dataset}
    \label{fig:graph}
\end{figure}

Firstly, the initial dataset of iterations is mapped to a social graph $\mathcal{G} = (\mathcal{N}, \mathcal{E})$ defined by a set of nodes $\mathcal{N}$ (users of the Twitter snapshot) and edges $\mathcal{E}$ (data flows between the users). As shown in \figureautorefname~\ref{fig:graph}, The social graph is directed and weighted, and the edge $(u,v,\omega)$ from user $u\in\mathcal{N}$ to user $v\in\mathcal{N}$ indicates that user $u$ has retweeted user $v$ in $\omega$ different times ($\exists \ t \ | \ u = t^{retweeter} \ \& \ v=t^{retweeter}$).

\begin{figure}[ht]
    \centering
    \includegraphics[width=\columnwidth]{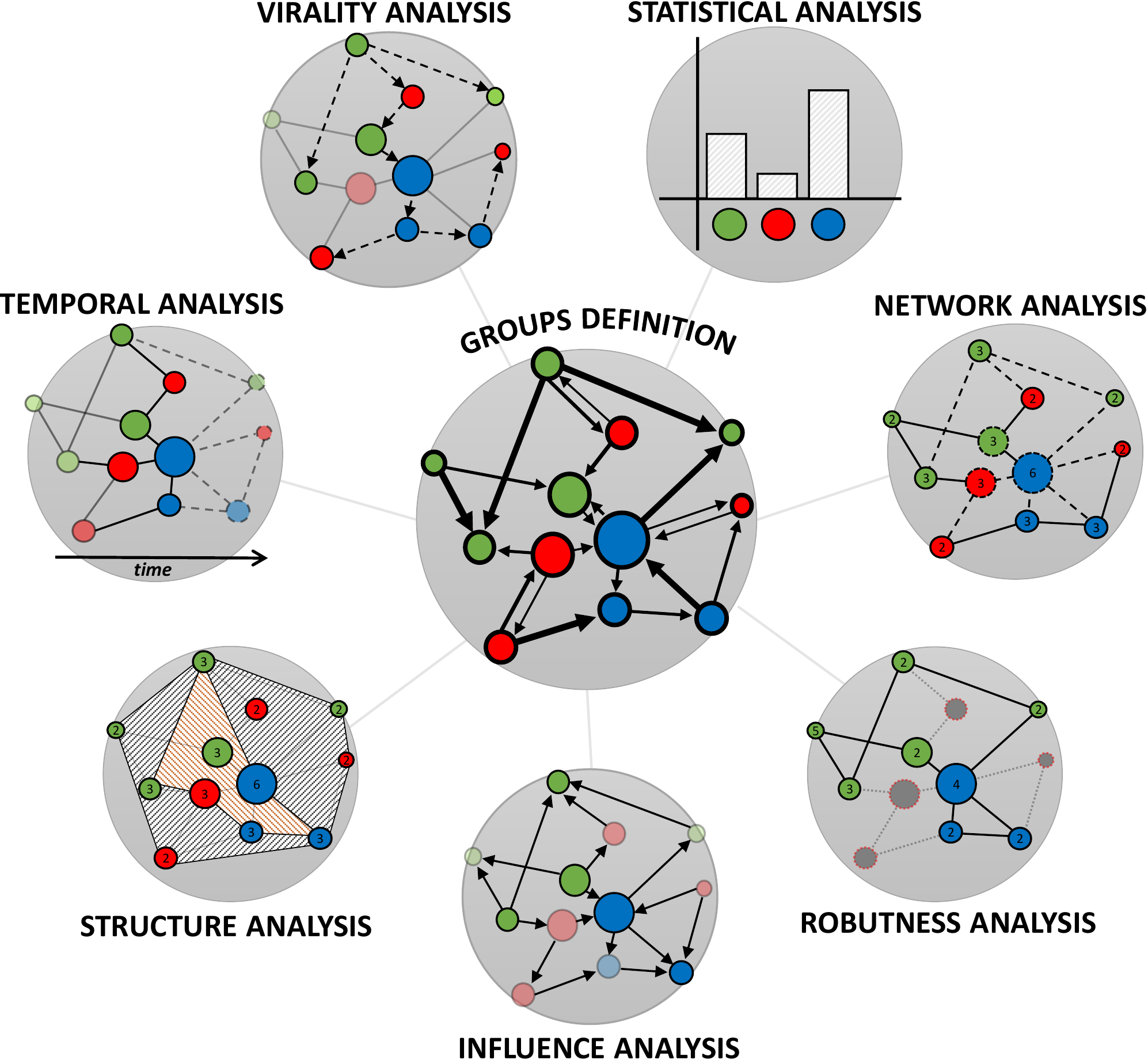}
    \caption{Analysis perspectives of BOTTER}
    \label{fig:framework}
\end{figure}

Secondly, given the graph $\mathcal{G}$, we are interested in comparing the statistical, network, robustness, influence, structure, temporal, content, and virality impact of specific subsets of users $\mathcal{N}_{1},\mathcal{N}_{2},\ldots, \mathcal{N}_{n} \subset \mathcal{N}$. %Note that these groupings of users are actually built by the criteria that is the reason for study. To mention some use cases, the framework could evaluate social bots interference in legitimate communications, compare the differences between political right wing, center, or left wing accounts, or characterize influencers with respect to the rest of users. 
\figureautorefname~\ref{fig:framework} shows the perspectives of the framework. As mentioned, we assume the most general and complete case of a directed and weighted social graph, although the framework would also be suitable for undirected and unweighted alternatives. In the following, the overall methodology is in-depth discussed.

%% file: framework/1-groups.tex
\subsection{Groups definition}
\label{sec:a}

\noindent \textbf{Goal}: Map each user within a specific cluster or group\\
\textbf{Research question}: How is the social graph grouped or divided?\\
\textbf{Description}: Users belonging to $\mathcal{N}$ should be mapped to their respective groups of study $\mathcal{N}_{1},\mathcal{N}_{2},\ldots, \mathcal{N}_{n}$ (Algorithm~\ref{alg:1}). If users are somehow grouped by predefined (finite) categories, we jump to Step 2 (see Section~\ref{sec:b}). On the contrary, if users are not categorized beforehand, groups are created prior to the execution of the analysis perspectives.

Therefore, each user $u$ is characterized through a \textit{categorization function} $f(u)$ of free choice. The latter should return the group of the user, $u^\text{group}$, that could be a categorical value (label) or a continuous/discrete value (number). For example, the categorization could be performed by a community detection algorithm that distinguishes suspicious actors from legitimate users, a machine learning classifier that returns the adherence to different ideologies, or a model that outputs the probability of being a social bot.

If the \textit{categorization function} directly returns the categories of users, sets $\mathcal{N}_{1},\mathcal{N}_{2},\ldots, \mathcal{N}_{n}$ will emerge. However, if a numeric value is assigned, analysts will finally map numerical ranges to categories. If no heuristic procedure is available, we suggest splitting the population with $n-1$ percentiles to define $n$ groups.

From the categorization of each account, the following comparative steps rely on analyzing the differences between the defined groups.\\

\begin{algorithm}[ht]
\SetAlgoLined
\KwData{\{$u \ | \ u \in \mathcal{N}$, and \! $u$ \! is \! not \! $categorized$\}}
\KwResult{$\mathcal{N}_{1},\mathcal{N}_{2},\ldots, \mathcal{N}_{n}$, \{$u^\text{group} \ | \ u \in \mathcal{N}$\}}
\texttt{\\}
$\mathcal{N}_{1},\mathcal{N}_{2},\ldots, \mathcal{N}_{n} \gets \emptyset$\;
 \For{$u \in \mathcal{N}$}{
  $c \gets f(u)$\;
  \uIf{$c$ is $categorical$}{
   $u^\text{group} \gets c$\;
   add $u$ to the subset $\mathcal{N}_{s} \subset \mathcal{N}$ associated with $c$\;
   }
   \ElseIf{$c$ is $numerical$}{
   $h \gets heuristic(c)$\;
   $u^\text{group} \gets h$\;
   add $u$ to the subset $\mathcal{N}_{s} \subset \mathcal{N}$ associated with $h$\;
  }
 }
\caption{Groups definition}\label{alg:1}
\end{algorithm}

%% file: framework/2-distribution.tex
\subsection{Statistical analysis}
\label{sec:b}
\noindent \textbf{Goal}: Quantify the number of users per group.\\
\textbf{Research question}: Is the social graph \textit{equally} or \textit{unevenly distributed}? Which are the \textit{majority group} or \textit{minority cluster}?\\
\textbf{Description}: Firstly, it is worth studying the distribution of users concerning their categorization (Algorithm~\ref{alg:2}). This exploration unveils whether groups are \textit{equally distributed} or \textit{unevenly distributed}. In other words, if the \textit{categorization function} $f(u)$ emits values following a uniform distribution or a non-uniform distribution.

An \textit{equally distributed} social graph reveals that users are differentiated in compensated communities. On the contrary, an \textit{unevenly distributed} ecosystem exposes unbalanced categories in which majority groups, minority clusters, and residual users can be identified. Tiny clusters could suggest anomalous or suspicious groupings.

In any case, these distribution results would not indicate a priori a greater or lesser impact of the groups on the ecosystem. However, they serve to contextualize and relativize the measurements of the next steps.

\begin{algorithm}[ht]
\SetAlgoLined
\KwData{\{$u^\text{group} \ | \ u \in \mathcal{N}$\}}
\KwResult{$|\mathcal{N}_{1}|,|\mathcal{N}_{2}|,\ldots,|\mathcal{N}_{n}|$}
\texttt{\\}
calculate cardinalities $|\mathcal{N}_{1}|,|\mathcal{N}_{2}|,\ldots,|\mathcal{N}_{n}|$\;

\eIf{$|\mathcal{N}_{1}|\approx|\mathcal{N}_{2}|\approx\ldots\approx|\mathcal{N}_{n}|$}{
$\mathcal{G}$ is \textit{equally distributed}\;
}{
$\mathcal{G}$ is \textit{unevenly distributed}\;
}
\caption{Statistical analysis}\label{alg:2}
\end{algorithm}

%% file: framework/3-composition.tex
\subsection{Network analysis}
\label{sec:c}
\noindent \textbf{Goal:} Explore network properties and characterize the groups under study.\\
\textbf{Research question}: How active, popular, consumer, producer, accessible, and diffuser are the groups?\\
\textbf{Description:} The situation being analyzed may vary from small networks to real-world and large networks. In this sense, the most complete the social graph is, the most reality it reflects. Nevertheless, analysts should consider the trade-off between representativeness and computational complexity. In any case, it is crucial to determine the magnitude and complexity of the graph $\mathcal{G} = (\mathcal{N}, \mathcal{E})$ through its main characteristics:
 
\begin{itemize}
    \item \textit{Graph order}, $|\mathcal{N}|$: Number of nodes/vertices of the graph, representing the total bulk of users involved in the social network snapshot.
    \item \textit{Graph size}, $|\mathcal{E}|$: Number of edges/links of the graph, which reveals the quantity of user-to-user relationships compiled in the experiment.
    \item \textit{Graph density/sparsity}, $d(\mathcal{G}) = \frac{|\mathcal{E}|}{ |\mathcal{N}|\cdot(|\mathcal{N}|-1)}$: Fraction of connected pairs of nodes for the maximum number of possible connections. If $d(\mathcal{G}) \ll 1$ and $|\mathcal{E}| \propto |\mathcal{N}|$, then the network is \textit{sparse}. On the contrary, if $d(\mathcal{G}) \sim 1$ and $|\mathcal{E}| \gg |\mathcal{N}|$, then the network is \textit{dense}.
\end{itemize}

Apart from the aforementioned superficial definitions, we can delve into other network properties to reveal specific insights about the network structure of social relationships and the dynamics of the information spreading phenomena. Putting the focus on each node $u$, we highlight the most generic, straightforward, and used attributes:

\begin{itemize}
    \item \textit{Node degree, $deg(u)=|neigh(u)|$}: Total number of edges or neighbors, which exposes the number of social links or data-flow paths in which user $u$ is involved. Additionally, if $\mathcal{G}$ is directed:
    
    \begin{itemize}
        \item \textit{Node in-degree, $deg_{in}(u)=|neigh_{in}(u)|$}: Number of incoming edges, which indicates social followers or content consumption trajectories to $u$. From a social perspective, this metric could be a synonym for \textit{activity}.
        
        \item \textit{Node out-degree, $deg_{out}(u)=|neigh_{out}(u)|$}: Number of outcoming edges, which represents social followings or content spread trajectories from $u$. From a social perspective, this metric could be a synonym for \textit{popularity}.
    \end{itemize}
    
    \item \textit{Node strength, $str(u)=\sum_{v \in neigh(u)}\omega(u,v)$}: Only if $\mathcal{G}$ is weighted, this attribute is computed as the sum of the weights of the edges, which states the total number of social interactions where $u$ is involved. Additionally, if $\mathcal{G}$ is directed:
    
    \begin{itemize}
        \item \textit{Node in-strength, $str_{in}(u)=\sum_{v \in neigh_{in}(u)}\omega(v,u)$}: Sum of the weights of the incoming edges, which reveals the number of messages shared by $u$. From a social perspective, this metric could be a synonym for content \textit{consumption}.
        
        \item \textit{Node out-strength, $str_{out}(u)=\sum_{v \in neigh_{out}(u)}\omega(u,v)$}: Sum of the weights of the outcoming edges, which reveals the times that $u$'s content has been disseminated. From a social point of view, this metric could be a synonym for content \textit{production}.
    \end{itemize}
    
    \item \textit{Node closeness/farness, $C(u)=\sum_{v\in\mathcal{N}}d(u,v)$}: Sum of the distances through shortest paths from $u$ to all others nodes. Therefore, the lowest values would suggest that these users are the closest to the rest of nodes, whereas the highest values would mean that these users are the most distant otherwise. Note that the inverse of that sum is known as \textit{closeness centrality}. From a social perspective, closeness could be a synonym for \textit{visibility}. 
    
    \item \textit{Node betweenness, $B(u)=\sum_{v,w\in\mathcal{N} | u \ne v \ne w} \frac{\sigma_{vw}(u)}{\sigma_{vw}}$}: Given the number of shortest paths from node $v$ to node $w$ ($\sigma_{vw}$), and those within the latter in which node $u$ is crossed ($\sigma_{vw}(u)$), this measurement represents the total number of times that node $u$ is involved in the total number of shortest paths of the graph. In this sense, the higher the count, the more frequent the user is involved in social dynamics such as diffusion cascades. From a social point of view, the betweenness could be understood as a synonym for \textit{diffusion}.

\end{itemize}

Note that there is a high amount of other topology-based measurements in order to explore more details about users and their interactions (distance, diameter, modularity, reciprocity, homophily, similarity, clustering coefficient, etc.~\cite{Pierri2020}), but they are out of scope of this framework. Nevertheless, analysts using this methodology can implement additional metrics to study patterns not covered by the attributes above.

Thus, the network composition can be implemented from a two-fold perspective.

\subsubsection{Global perspective}

The objective is to measure the changes of the network topology as we incrementally insert the nodes and edges of the defined groups to finally get the current snapshot under study (Algorithm~\ref{alg:3}). For example, to evaluate the interference of malicious spreaders, we monitor the changes in network metrics along with the social graph as we add new nodes and edges in order of ``malice". The latter permits a high-level evaluation of the global network alteration while we include the illegitimate actors.

We start from the defined groups ($\mathcal{N}_{1},\mathcal{N}_{2},\ldots,\mathcal{N}_{n}$) and respective edges ($\mathcal{E}_{1},\mathcal{E}_{2},\ldots,\mathcal{E}_{n}$). Therefore, network properties (order, size, and density/sparsity) and aggregated node attributes (average degree, average strength, average closeness/farness, average betweenness) are computed for the  incrementally built graph $\mathcal{G}_{Acum \cup g}$. The latter is formed explicitly through the steps:

\noindent$\begin{array}{lclcl}

\mathcal{G}_{Acum \cup g}(1) & = & \mathcal{G}_{1} & = &  (\mathcal{N}_{1}, \mathcal{E}_{1}),\\

\mathcal{G}_{Acum \cup g}(2) & = & \mathcal{G}_{1 \cup 2} & = & (\mathcal{N}_{1}\!\cup\!\mathcal{N}_{2}, \mathcal{E}_{1}\!\cup\! \mathcal{E}_{2}),\\

\mathcal{G}_{Acum \cup g}(n) & = & \mathcal{G}_{1 \cup \cdots n} & = & (\mathcal{N}_{1}\!\cup\!\cdots\mathcal{N}_{n},\ \mathcal{E}_{1}\!\cup\!\cdots\mathcal{E}_{n})

\end{array}$

Then, if network properties and averaged node attributes remain proportional as we incrementally build the social graph $\mathcal{G}_{Acum \cup g}$, groups $\mathcal{N}_{1},\mathcal{N}_{2},\ldots,\mathcal{N}_{n}$ are suggested to constitute an \textit{ecosystem maintainer}. On the contrary, if the inclusion of a new group distorts metrics, that group is supposed to be an \textit{ecosystem changer} that impacts the social network.

\begin{algorithm}[ht]
\SetAlgoLined
\KwData{$\mathcal{G} = (\mathcal{N}, \mathcal{E})$}
\KwResult{Properties of $\mathcal{G}_{1}, \mathcal{G}_{1 \cup 2}, \ldots, \mathcal{G}_{1 \cup 2 \cup \ldots \cup n}$}
\texttt{\\}

$\mathcal{N}_{Acum}, \mathcal{E}_{Acum}, \mathcal{G}_{Acum} \gets \emptyset$\;

 \For{$g\in \{1, 2, 3,\ldots,n\}$}{
  $\mathcal{N}_{Acum \cup g} \gets \mathcal{N}_{Acum} \cup \mathcal{N}_{g}$\;
  $\mathcal{E}_{Acum \cup g} \gets \mathcal{E}_{Acum} \cup \mathcal{E}_{g}$\;
  $\mathcal{G}_{Acum \cup g} \gets (\mathcal{N}_{Acum},\mathcal{E}_{Acum})$\;

  calculate $|\mathcal{N}_{Acum \cup g}|, |\mathcal{E}_{Acum \cup g}|$ and $d(\mathcal{G}_{Acum \cup g})$\;
  
  %save network properties of $\mathcal{G}_{Acum\cup g}$\;
  
   \For{$u\in \mathcal{N}_{Acum}$}{
     calculate $deg_{in}(u), deg_{out}(u), str_{in}(u), str_{out}(u), C(u), B(u)$;
   }
   %save averaged node attributes of $\mathcal{G}_{Acum \cup g}$\;
   $\mathcal{N}_{Acum} \gets \mathcal{N}_{Acum \cup g} $\;
   $\mathcal{E}_{Acum} \gets \mathcal{E}_{Acum \cup g} $\;
   $\mathcal{G}_{Acum} \gets \mathcal{G}_{Acum \cup g} $\;
   
 }
   \eIf{network properties of $\mathcal{G}_{1} \propto \mathcal{G}_{1 \cup 2} \propto \ldots \propto \mathcal{G}_{1 \cup 2\cup\ldots\cup n}$ 
   or averaged node attributes of  $\mathcal{G}_{1} \simeq \mathcal{G}_{1 \cup 2}\simeq \ldots \simeq \mathcal{G}_{1 \cup 2\cup\ldots\cup n}$ }{
    groups $\mathcal{N}_{1},\mathcal{N}_{2},\ldots,\mathcal{N}_{n}$ are \textit{ecosystem maintainers}\;
   }{
    some groups $\mathcal{N}_{1},\mathcal{N}_{2},\ldots,\mathcal{N}_{n}$ are \textit{ecosystem changers}\;
  }
\caption{Global network composition}\label{alg:3}
\end{algorithm}

\subsubsection{Nodes perspective}

The objective is to characterize the defined groups through their node attributes. In this case, we do not directly measure the impact of the groups in the global network (Algorithm~\ref{alg:4}). Nonetheless, group features differences detected in this perspective will be correlated with distortions in network properties. This independent analysis can also be performed to understand group differences detected in the global perspective.

Therefore, the node attributes (degree, strength, closeness/farness, betweenness) are calculated and averaged for each group ($\mathcal{N}_{1},\mathcal{N}_{2},\ldots,\mathcal{N}_{n}$). These measurements enable a direct comparison between groups through their corresponding nodes and edges. If groups $\mathcal{N}_{1},\mathcal{N}_{2},\ldots,\mathcal{N}_{n}$ are considered \textit{ecosystem maintainers} from a global perspective, then feature distributions will be uniform from this node perspective, indicating they \textit{behave similarly}. Analogously, groups constituting \textit{ecosystem changers} that cause an impact in the social graph will have particular patterns from the node perspective, meaning that they \textit{behave differently}.

\begin{algorithm}[ht]
\SetAlgoLined
\KwData{$\mathcal{G} = (\mathcal{N}, \mathcal{E})$}
\KwResult{Properties of $\mathcal{G}_{1}, \mathcal{G}_{2},\ldots, \mathcal{G}_{n}$}
\texttt{\\}

 \For{$g\in \{1,2,3,\ldots,n\}$}{

   \For{$u\in \mathcal{N}_{g}$}{
     calculate $deg_{in}(u), deg_{out}(u), str_{in}(u), str_{out}(u), C(u), B(u)$;
   }
   %save averaged node attributes of $\mathcal{G}_{g}$\;
 }
   \eIf{averaged node attributes of  $\mathcal{G}_{1} \simeq \mathcal{G}_{2}\simeq \ldots \simeq \mathcal{G}_{n}$ }{
    groups $\mathcal{N}_{1},\mathcal{N}_{2},\ldots,\mathcal{N}_{n}$ \textit{behave similarly}\;
   }{
    groups $\mathcal{N}_{1},\mathcal{N}_{2},\ldots,\mathcal{N}_{n}$ \textit{behave differently}\;
  }
\caption{Node network composition}\label{alg:4}
\end{algorithm}

%% file: framework/4-robustness.tex
\subsection{Robustness analysis}
\label{sec:d}

\noindent \textbf{Goal:} Measure vulnerability of the social graph through
the disconnection of nodes and detect critical groups.\\
\textbf{Research question:}  Do social dynamics change when nodes are removed? Do malicious groups interfere in the network?\\
\textbf{Description:} A system is considered robust if the failure of some of its components does not negatively impact the rest. This perspective aims to measure the robustness of the social graph by checking the effect of node deletions in the connectivity of the network and the number of total edges and weight. 

Intuitively, we assume that the social graph is robust if the number of remaining users after removing is proportional to the total weights, total edges, and the number of nodes in the giant component (the subgraph that connects the largest number of nodes, leaving out those that are not connected to it) of the resulting social graph.

The proposed robustness analysis (Algorithm~\ref{alg:5}) consists of calculating the variation of the network properties total weight, total edges, and size of the giant component as we iteratively disconnect groups $\mathcal{N}_{1},\mathcal{N}_{2},\ldots,\mathcal{N}_{n} \in \mathcal{N}$. We work with an extra copy of the graph, $\mathcal{G}'$, that saves the remaining social graph after the group eliminations. 

First, we calculate the fraction of users representing the group $g$ within the total number of users ($p_{g}=|\mathcal{N}_{g}| / |\mathcal{N}|$). Therefore, we infer the associated theoretical network properties ($total\_edges_{g}$, $total\_weight_{g}$, and $g\_comps_{g}$), proportional to the fraction $p_{g}$ and serving as a baseline for the robustness analysis of group $g$. For example, if the first group covers half (0.5) of the total number of users, the proportional reference would be that they are responsible for the half of total edges, weights, and the number of nodes in the giant component.

Second, we propose to inspect a group $g$ in different steps by removing portions of users ($r$, from 0.2 of the group to the whole group) instead of just eliminating it completely. On the one hand, in each cumulative removal, we calculate the associated total weight, total edges, and size of the giant component ($total\_edges_{g,r}$, $total\_weight_{g,r}$, and $g\_comps_{g,r}$) in the resulting social graph $\mathcal{G}'_{g,r}$ (the one composed by the $1-r$ fraction of $g$ and remaining groups that have not yet been analyzed). On the other hand, we compute a theoretical baseline that assumes that network properties of the resulting social graph $\mathcal{G}'_{g,r}$ gets reduced in proportion to the number of removed users ($(1-r)*total\_edges_{g}$, $(1-r)*total\_weight_{g}$, and $(1-r)*g\_comps_{g}$). Following the example, if we remove two out of ten users (0.2) of the first group (that represented half of the total), the baseline would be that the remaining users are responsible for the 0.1 (0.5 * 0.2) of total edges, weights, and the number of nodes in the giant component.

To evaluate the robustness of the network to each removal, we compare the practical, measured values ($total\_edges_{g,r}$, $total\_weight_{g,r}$, and $g\_comps_{g,r}$) with the proportional estimations ($(1-r)*total\_edges_{g}$, $(1-r)*total\_weight_{g}$, and $(1-r)*g\_comps_{g}$). Such subsamples of $g$ that experiment similar quantities suggest that they have a expected impact, but differences indicate disproportional decrements in network properties and expose outstanding or anomalous users. In other words, a fraction $r$ of users belonging to a group $g$ provoking non-proportional damage in the topology much greater than $r*p_{g}$ could be considered as \textit{destabilizing} users (imagine the example in which the removed users of the first group impacts in a reduction larger than 0.1 in network properties). On the contrary, those disconnected groups without extraordinary transcendence could be considered \textit{non-destabilizing} ones (in the running example, the first removal of the first group reduces metrics by 10\% or less).

%Once a group has been analyzed along with different removal simulations, then the social graph $\mathcal{G}'$ is updated by removing the group \textit{g} ($\mathcal{G}_{g}$).

\begin{algorithm}[ht]
\SetAlgoLined
\KwData{$\mathcal{G} = (\mathcal{N}, \mathcal{E})$}
\KwResult{Network robustness against $\mathcal{N}_{1},\mathcal{N}_{2},\ldots,\mathcal{N}_{n}$}
\texttt{\\}

$total\_edges \gets |\mathcal{E}|$\;
$total\_weight \gets \omega(\mathcal{E})$\;
$g\_comps \gets giant\_component(\mathcal{G})$\;
$\mathcal{G}' \gets \mathcal{G}$\;

 \For{$g\in \{1,2,3,\ldots,n\}$}{

    $ p_{g} \gets |\mathcal{N}_{g}| / |\mathcal{N}| $\;
    $total\_edges_{g} \gets p_{g}*total\_edges$\;
    $total\_weight_{g} \gets p_{g}*total\_weight$\;
    $g\_comps_{g} \gets p_{g}*g\_comps$\;

   \For{$r \in \{0.2, 0.4, 0.6, 0.8, 1.0\}$}{
   
     $\mathcal{G}'_{g,r} \gets $ remove fraction $r$ of $\mathcal{N}_{g}$ from $\mathcal{G}'$\;
     $total\_edges_{g,r} \gets |\mathcal{E}'_{g,r}|$\;
     $total\_weight_{g,r} \gets \omega(\mathcal{E}'_{g,r})$\;
     $g\_comps_{g,r} \gets giant\_component(\mathcal{G}'_{g,r})$\;
    
        \eIf{$total\_edges_{g,r} \approx (1-r)*total\_edges_{g}$ and $total\_weight_{g,r} \approx (1-r)*total\_weight_{g}$ and $g\_comps_{g,r} \approx (1-r)*g\_comps_{g}$}{
            a portion $r$ of $\mathcal{G}_g$ are \textit{non-destabilizing} users\;
         }{
            a portion $r$ of $\mathcal{G}_g$ are \textit{destabilizing} users\;
        }
  }
  $\mathcal{G}' \gets \mathcal{G}-\mathcal{G}_{g}$
}
\caption{Robustness analysis}
\label{alg:5}
\end{algorithm}

%% file: framework/5-spreading.tex
\subsection{Influence analysis}
\label{sec:e}
\noindent \textbf{Goal}: Detect and study the role of the groups in information spreading.\\
\textbf{Research question}: Which groups are producers, consumers, and spreaders of information?\\
\textbf{Description}: The propagation of messages through social networks has been studied for years. In this sense, the community has proposed several metrics to characterize the role and importance of individual users in such a phenomenon~\cite{Zheng2014}. The most renowned algorithms are as follows:

\begin{itemize}
    \item \textit{PageRank}, $PR(u)=\frac{1-d}{|\mathcal{N}|} + d * \sum_{v \in neigh_{in}(u)}{\frac{PR(v)*\omega(u,v)}{deg_{out}(v)}}$, where $d$ is a damping factor, frequently 0.85. This rating score reveals the relative importance of user $u$ in the social graph given the quantity and quality of the incoming edges. This centrality measure, which recursively depends on the sum of the ranks of the in-neighbors, is larger when the social environment is highly viral. Although Google designed it for ranking websites, it has been widely adopted for social media studies~\cite{Riquelme2016}.
    
    \item \textit{Hyperlink-Induced Topic Search}, $HITS_{H}(u) = \sum_{v \in neigh_{out}(u)}{\omega(u,v)*HITS_{A}(v)}$, $HITS_{A}(u) = \sum_{v \in neigh_{in}(u)}{\omega(v,u)*HITS_{H}(v)}$. This link analysis algorithm  recursively calculates both hub and authority values for node $u$. In an information spreading network that traces the flows of posts, the former expresses the capacity of $u$ to share content. In contrast, the latter represents the consumption and dissemination of organic content by $u$. Thus, a powerful hub will link to excellent authorities, and relevant authorities will be referenced by many important hubs. Originally, this algorithm was also developed for rating web pages, but its usage has been expanded to other networking use cases~\cite{Riquelme2016}.
    
   \item \textit{Eigenvector centrality}, $eig(u)= \frac{1}{\lambda} \sum_{v \in neigh(u)} \, eig(v)$, where $\lambda \neq 0$ is a constant known as eigenvalue. This centrality measure is widely used to measure the global influence of a node~\cite{Riquelme2016}, considering the importance of its neighbors. Node $u$ will have a more powerful eigenvector centrality if it is surrounded by nodes with also high values. The latter would suggest that $u$ is located in an area with a relevant exchange of information and messages (in a retweet network).
\end{itemize}

In this analysis perspective, as observed in Algorithm~\ref{alg:6}, it is worth unveiling relationships between the defined groups and the metrics above. Therefore, we should calculate them for each node of the snapshot ($PageRank(u), HITS_{H}(u), HITS_{A}(u), eig(u)$) and average per group ($\mathcal{N}_{1}, \mathcal{N}_{2},\ldots, \mathcal{N}_{n}$). Generally, having common averaged results would mean that all groups \textit{influence similarly}. On the contrary, when some groups stand out from the rest in such metrics, it is assumed that groups \textit{influence differently}. 

\begin{algorithm}[t]
\SetAlgoLined
\KwData{$\mathcal{G} = (\mathcal{N}, \mathcal{E});$}
\KwResult{Network influence of $\mathcal{N}_{1}, \mathcal{N}_{2},\ldots, \mathcal{N}_{n}$}
\texttt{\\}

 \For{$g\in \{1,2,3,\ldots,n\}$}{

   \For{$u\in \mathcal{N}_{g}$}{
     calculate $PageRank(u), HITS_{H}(u), HITS_{A}(u), eig(u)$;
   }
   %save averaged spreading metrics of $\mathcal{G}_{g}$\;
 }
   \eIf{averaged influence of  $\mathcal{N}_{1} \simeq \mathcal{N}_{2}\simeq \ldots \simeq \mathcal{N}_{n}$ }{
    groups $\mathcal{N}_{1},\mathcal{N}_{2},\ldots,\mathcal{N}_{n}$ \textit{influence similarly}\;
   }{
    groups $\mathcal{N}_{1},\mathcal{N}_{2},\ldots,\mathcal{N}_{n}$ \textit{influence differently}\;
  }
\caption{Influence analysis}\label{alg:6}
\end{algorithm}

%% file: framework/6-slicing.tex
\subsection{Structure analysis}
\label{sec:f}
\noindent \textbf{Goal}: Partition the graph $\mathcal{G}$ in $k$ layers to study contact networks and data paths in which users are involved.\\
\textbf{Research questions}: Are groups different in the number of contacted users? Do malign nodes participate in numerous data flows?\\
\textbf{Description}: Traditionally, the structure of social networks has been explained with the $k$-shell decomposition~\cite{Zheng2014}. This method partitions the graph into layers according to connectivity patterns and assigns an integer $k$ to each node. In particular, the layers range from the outer layer $k=1$ to the inner layer $k=max$ and group the nodes with the same degree \textit{k}. Therefore, nodes with degree $k$ form the $k$-shell. %Additionally, the $k$-core of a graph $\mathcal{G}$ is the union of shells from $k$-shell to $k_{max}$-shell. The degree of the nodes in a $k$-core is, at least, $k$.  

In an information spreading network, such sub-partitions enable the analysis of spreading zones and interaction spheres as we get closer to the core. A user with a high $k$ is very present in data transmission paths and highly connected to other nodes of the core. On the contrary, a user with a low value of $k$ is situated in the periphery, meaning a poor interaction in social network trends.

In this way, as reflected in Algorithm~\ref{alg:7}, we can study whether each group (${\mathcal{N}_{1},\mathcal{N}_{2},\mathcal{N}_{3},\ldots,\mathcal{N}_{n}}$) maintains its portion of representation ($p_g$) in $\mathcal{N}$ across the different $k$-shells. In this sense, an equal portion of $g$ in the different $k$-shells ($p_{g,k}$) entails that members are \textit{proportionately located} in different areas of propagation of the social network. On the contrary, we could detect anomalies in a snapshot where groups are \textit{disproportionately located} when $k$-shells do not follow a fair representation of each group and layers are formed only by members of certain groups. For example, if the inner $k$-shells are \textit{highly populated} by a specific group, this would indicate that the core of the social network is highly biased towards that group. Accordingly, the outer $k$-shells will be \textit{quite depopulated} of that group of users. On the other hand, a group mainly present in the outer $k$-shells is doomed not no reach the rest of the users.

\begin{algorithm}[ht]
\SetAlgoLined
\KwData{$\mathcal{G} = (\mathcal{N}, \mathcal{E})$}
\KwResult{Network locations of $\mathcal{N}_{1},\mathcal{N}_{2},\ldots,\mathcal{N}_{n}$}
\texttt{\\}

$k$-shells $\gets$ calculate $k$-shells of $\mathcal{G}$

 \For{$g\in \{1,2,3,\ldots,n\}$}{

    $ p_{g} \gets |\mathcal{N}_{g}| / |\mathcal{N}| $\;
    
   \For{$k \gets 1$ to $max$}{
   
     $ p_{g,k} \gets \frac{|\{ u \ | \  u \in \mathcal{N}_{g} \ \wedge \ u \in kshell \}|}{|\{ u \ | \ u \in kshell \}|} $
    
    \eIf{$ p_{g,k} \approx p_{g} $}{
        $g$ is \textit{proportionately located} in $k$-shell\;
     }{
        $g$ is \textit{disproportionately located} in $k$-shell\;
        \eIf{$ p_{g,k} > p_{g} $}{
            $k$-shell is \textit{highly populated} by $g$\;
        }{
            $k$-shell is \textit{quite depopulated} of $g$\;
        }
    }
  }
}
\caption{Structure analysis}
\label{alg:7}
\end{algorithm}

%% file: framework/7-timeline.tex
\subsection{Temporal analysis}
\label{sec:g}
\noindent \textbf{Goal}: Analyze time series to reveal. temporal patterns or anomalies and categorize the activity of groups over time.\\
\textbf{Research questions}: What is the proportion of created content over time? Is there any suspicious traffic at specific periods? Are there groups stimulated in particular situations? \\
\textbf{Description:} All algorithms mentioned above assume a static scenario where the bulk of interactions and users are analyzed without considering their temporality. However, social networks are naturally the reflection of day-to-day life, and the characteristics of the social graph may vary accordingly. Therefore, the extraction of temporal patterns may help understand several phenomena of the real-life~\cite{Alizadeh2020}. 

In Algorithm~\ref{alg:8}, we propose to measure the impact of each group $\mathcal{N}_1,\mathcal{N}_2,\mathcal{N}_3,\ldots,\mathcal{N}_n$ in the ecosystem by quantifying the traffic emitted to the network per time unit (\eg days). We could assume that users of a certain group are \textit{normally stimulated} on a specific time unit if their interactions in that period ($p_{g,d}$) are proportional to the size of the group ($p_{g}$). Socially, this would mean that the events within such time unit have not specially activated these users. On the contrary, a group has been \textit{understimulated} in a time unit $d$ if its interactions ($p_{g,d}$) are less representative than expected because of their group size ($p_{g}$). Analogously, an extraordinary amount of traffic in specific periods would point to an \textit{overstimulated} attitude. Both situations could reveal relevant patterns, such as the level of interest in specific events, the sensitivity to particular situations, or the employment of targeting strategies. 

It is worth noting that detected anomalies in specific time units could be analyzed with the framework at hand with the subgraph $\mathcal{G}_{time\_units}=(\mathcal{N}_{time\_units}, \mathcal{E}_{time\_units})$ as an input. The latter would be composed only of the users who have interacted throughout those dates.

\begin{algorithm}[ht]
\SetAlgoLined
\KwData{$\mathcal{G} = (\mathcal{N}, \mathcal{E})$, Dataset of interactions}
\KwResult{Temporal activity of $\mathcal{N}_1,\mathcal{N}_2,\mathcal{N}_3,\ldots,\mathcal{N}_n$}
\texttt{\\}

 \For{$g\in \{1,2,3,\ldots,n\}$}{

    $ p_{g} \gets |\mathcal{N}_{g}| / |\mathcal{N}| $\;
    
   \For{$d \gets d_{1}$ to $d_{m}$}{
   
     $ p_{g,d} \gets \frac{|\{t \ | \ t\textsuperscript{retweeter}\in\mathcal{N}_g \ \wedge \ t\textsuperscript{timestamp} = d\}|}{|\{ t \ | \ t\textsuperscript{timestamp} = d \}|}$;
    
    \uIf{$ p_{g,d} \approx p_{g} $}{
        $g$ is \textit{normally stimulated} on $d$\;
     }
     \uElseIf{ $p_{g,d} < p_{g}$ }{
        $g$ is \textit{understimulated} on $d$\;
    }\Else{
        $g$ is \textit{overstimulated} on $d$\;
    }
  }
}
\caption{Temporal analysis}
\label{alg:8}
\end{algorithm}

%Several works have proposed sophisticated time series analysis and forecasting techniques to categorize social network dynamics [REF]. Unfortunately, the management of temporal data needs special efforts due to its complexity, so it is left out of this first version of the framework. We believe, though, that working in this line is definitely worthy to exploit the knowledge beyond time series.

%% file: framework/8-virality.tex
\subsection{Virality analysis}
\label{sec:h}
\noindent \textbf{Goal}: Trace viral content and analyze the speed of spreading.\\
\textbf{Research questions}: Who creates the most trending content? How are information cascades built over time?\\
\textbf{Description:}

The semantic of the content and the phenomenon of its propagation is relevant in social networks. Messages may be highly elaborated or cover relevant topics, but they will not reach a broad audience if they are not effectively disseminated. It is crucial the role of users sharing content with followers or friends on these platforms, deriving in an iterative process over time. In this sense, the study of information cascades reveals details such as the temporality in the propagation of content and the number of users involved~\cite{Pierri2020}.

Firstly, as seen in Algorithm~~\ref{alg:9}, we propose to explore which type of content is generated by each group $\mathcal{N}_1,\mathcal{N}_2,\mathcal{N}_3,\ldots,\mathcal{N}_n$. A good approximation is to extract \textit{word clouds} with the most common topics ($common\_topics_g$). If their themes are closely similar (from a semantic point of view), it would mean that groups \textit{discuss similarly} on the same topics. On the contrary, disparate \textit{word clouds} would demonstrate that such groups \textit{discuss differently} on various topics.

Secondly, we suggest studying the information cascades aggregating messages created per group (${t | t^{tweeter}\in\mathcal{G}_g}$) and extracting, from tweets ($t^{tweet}$), the size of provoked cascades (number of users sharing), and their duration (time over which messages continue to circulate without stagnating) to calculate their relationship ($cascades_g$).  Particularly, a group of \textit{influencers} is characterized by accumulating large cascades in short periods (e.g., it is common for their content to be massively shared within a few hours). On the other hand, a group of \textit{non-influencers} generates tiny cascades over an extensive range of times (e.g., their messages do not have a remarkable impact and may be sporadically shared throughout several days).

On average, if groups $\mathcal{N}_1,\mathcal{N}_2,\mathcal{N}_3,\ldots,\mathcal{N}_n$ produce cascades ($cascades_g$) with similar size and active time patterns, they could be considered \textit{equally viral}. However, if groups exhibit different relationships in both cascade sizes and duration, they are \textit{unevenly viral}.

\begin{algorithm}[ht]
\SetAlgoLined
\KwData{$\mathcal{G} = (\mathcal{N}, \mathcal{E})$, Dataset of interactions}
\KwResult{Content alignment and virality of $\mathcal{N}_1,\mathcal{N}_2,\ldots,\mathcal{N}_n$}
\texttt{\\}

 \For{$g\in \{1,2,3,\ldots,n\}$}{

   \For{$\{ t \ | \ t^{tweeter} \in \mathcal{G}_g \}$}{
     extract $t^{topics}$\;
     calculate cascade size and duration of $t^{tweet}$\;
   }
   $common\_topics_g \leftarrow$ most frequent topics of $g$\;
   $cascades_g \leftarrow$ averaged relations between cascade size and duration of $\mathcal{N}_g$\;
   
   \eIf{$cascades_g$ have big cascades in short periods}{
    $\mathcal{N}_g$ would be a group of \textit{influencers}\;
   }{
    $\mathcal{N}_g$ would be a group of \textit{non-influencers}\;
  }
   
 }
   \eIf{$common\_topics_1 \simeq common\_topics_2 \simeq \ldots \simeq common\_topics_n$ }{
    groups $\mathcal{N}_1,\mathcal{N}_2,\ldots,\mathcal{N}_n$ \textit{discuss similarly}\;
   }{
    groups $\mathcal{N}_1,\mathcal{N}_2,\ldots,\mathcal{N}_n$ \textit{discuss differently}\;
  }
  
     \eIf{$cascades_1 \simeq cascades_2 \simeq \ldots \simeq cascades_{n}$ }{
    groups $\mathcal{N}_1,\mathcal{N}_2,\ldots,\mathcal{N}_n$ are \textit{equally viral}\;
   }{
    groups $\mathcal{N}_1,\mathcal{N}_2,\ldots,\mathcal{N}_n$ are \textit{unevenly viral}\;
  }
\caption{Virality analysis}\label{alg:9}
\end{algorithm}

%% file: experiments.tex
\section{Experimental validation: the 2019 Spanish general election}\label{experiments}

We demonstrated the presence of social bots in Twitter political discussions in previous work~\cite{9226407}. In particular, we monitored Twitter for 38 days before the 2019 Spanish general election, November 11, discovering that malicious automated accounts strategically acted in key days, specially in the last week, by mainly spreading organic content to generate dispute between different political ideologies. However, we identified the challenge to measure the real impact of such fake users activity from such a scenario.

Therefore, with the proposed framework at hand, we are ready to evaluate whether such social bots really interfered in the normal Twitter dynamic. For this purpose, we collected tweets created between 2019-10-04 and 2019-11-11 containing at least one hashtag from a predefined list of 46 different politic-based hashtags~\cite{PASTORGALINDO2020106047}. On this occasion, we used the Enterprise version of the Twitter API instead of the Standard one to gather more data and cover a broader snapshot of the social network. As a result, we obtained a dataset with 2,802,467 users and 39,344,305 retweets. Nevertheless, to reduce the computational complexity of algorithm execution but simultaneously maintain a strong representation, we work with a random sample of the 20\% of the collected interactions, that is, 7,868,861 retweets 1,297,975 associated users.

In the following subsections, we launch the framework analysis perspectives against the explained use case. The Python code of the analysis is publicly available\footnote{github.com/javier-pg/botbusters-spanish-general-elections-network-analysis}. 

\subsection{Groups definition}

To measure the interference of social bots in our snapshot, we firstly evaluate each account with BotometerLite~\cite{Yang_Varol_Hui_Menczer_2020}, the de facto standard tool to calculate the automation (botscore) of a Twitter profile. In the line of Algorithm~\ref{alg:1}, our \textit{categorization function} is represented by $botscore(user)\rightarrow[0,1]$ and applied to the list of 1,297,975 sampled accounts. 

As the botscore of a Twitter account is a continuous value from 0.0 (no automation) to 1.0 (full automation), we adopt the heuristic of statically mapping these values to categorical labels. Several works perform binary classification of bot or not~\cite{9226407}, but we decide to study also the role of an intermediate band of semi-automated accounts. As there is not a homogeneous manner to map botscore to classes~\cite{https://doi.org/10.1002/hbe2.115}, we assume that 70\% with the lowest botscores are \humans\ (low automation), the 10\% with the highest botscores are \bots\ (high automation), and the remaining 20\% between the two groups are \semibots\ (intermediate automation). 

\begin{figure}[h!]
    \centering
    \begin{subfigure}[b]{\columnwidth}
        \centering
        \includegraphics[width=\columnwidth]{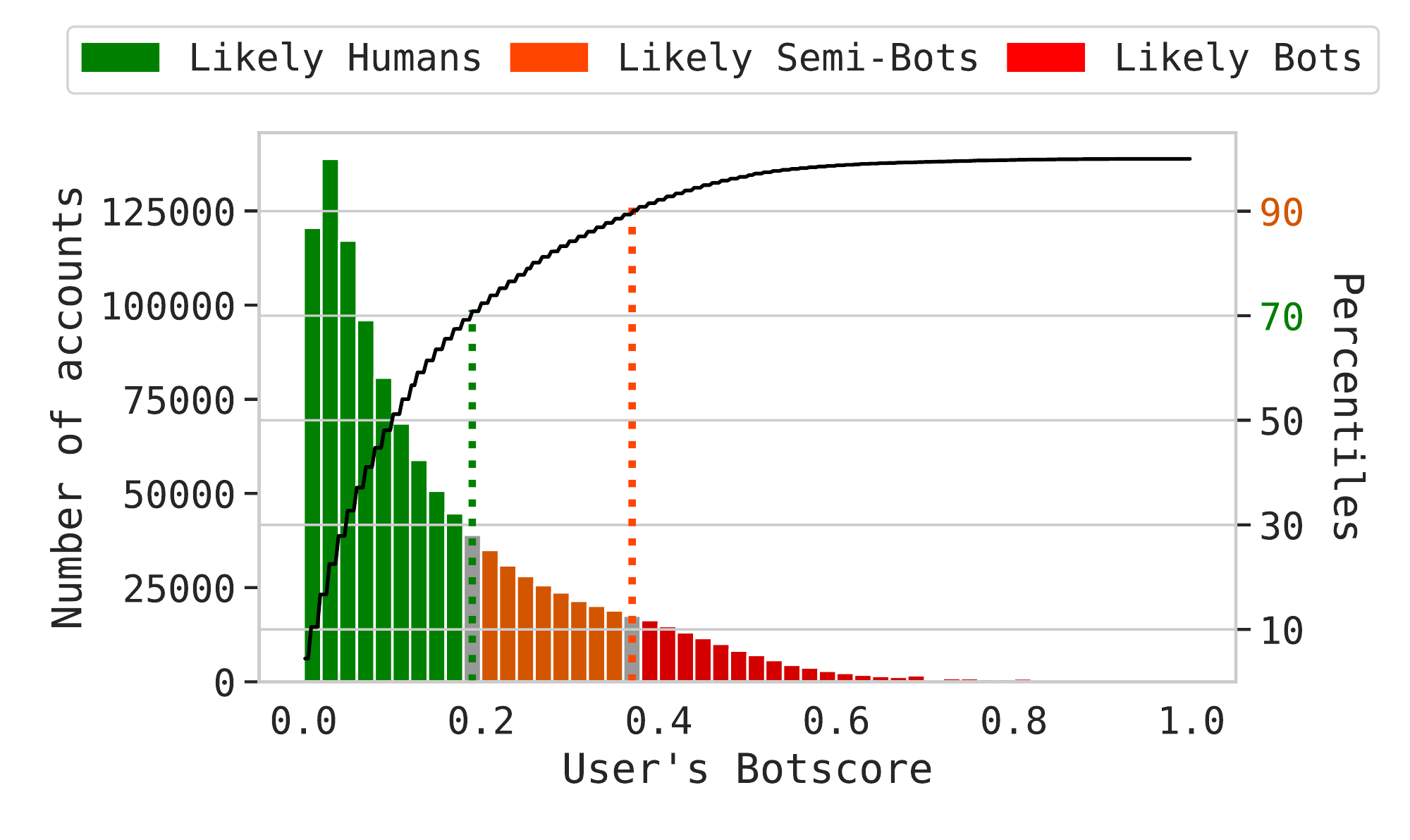}
        \caption{Botscore distribution and defined groups}
        \label{fig:distribution}
    \end{subfigure}
    \begin{subfigure}[b]{\columnwidth}
        \centering
        \includegraphics[width=0.9\columnwidth]{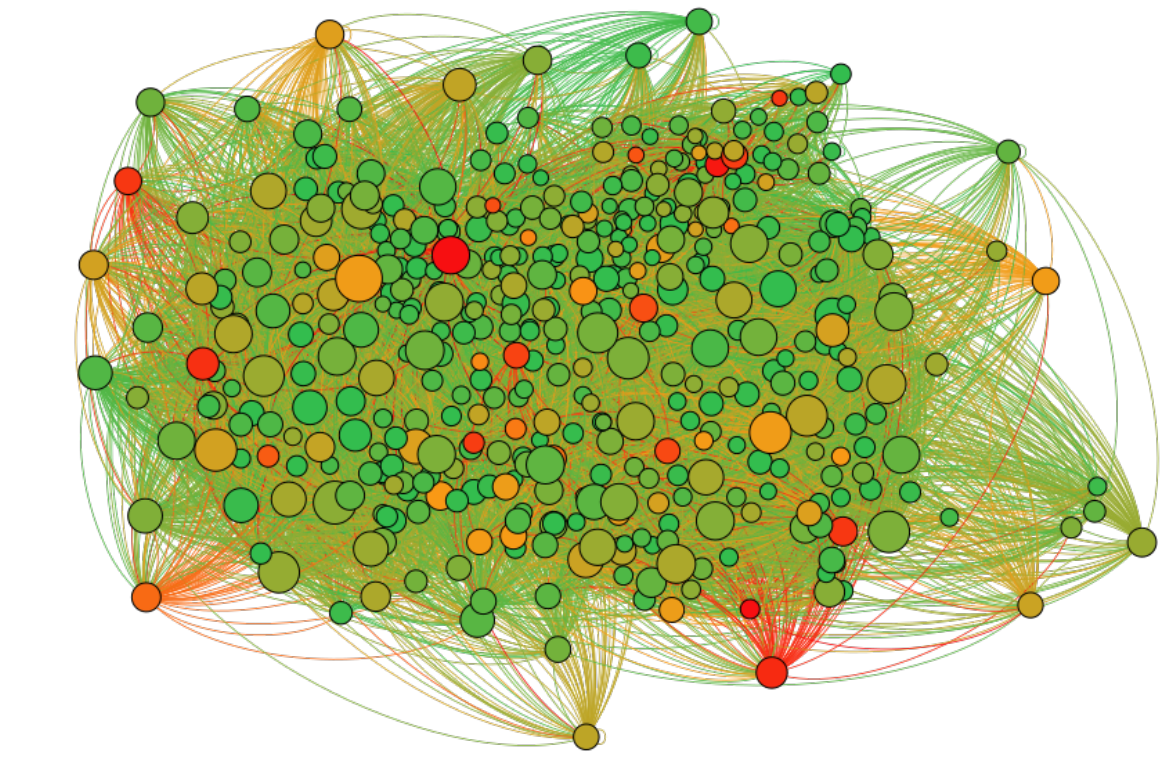}
        \caption{Giant component of nodes with more than degree 320}
        \label{fig:network}
    \end{subfigure}
\end{figure}

On the one hand, \figureautorefname~\ref{fig:distribution} shows the distribution (bar plot) and cumulative distribution (line plot) of the calculated botscores. We consider as \humans\ those legitimate users whose botscores are under the $70^{th}$ percentile (0.0$<$=botscore$<$0.19); as \semibots\ those accounts which implement some automatic tasks and have a botscore between 70$^{th}$ and 90$^{th}$ percentile (0.19$<$=botscore$<$0.37); and as \bots\ those fully-automated actors who have a boscore above the 90$^{th}$ percentile (0.37$<$=botscore$<$=1.0). On the other hand, \figureautorefname~\ref{fig:network} shows the giant component of the social graph composed of nodes with a degree greater than 320. The color of the nodes is associated with the class of the user (green corresponds to \humans, orange would be \semibots, and red represents \bots). At the same time, edges have the color of who performs the action (responsible for forwarding posts).

Despite defining three groups for better understanding the type of users beyond the calculated automation, we still use the continuous value of the botscore for analysis perspectives of the framework to be more precise than a ternary classification.

\subsection{Statistical analysis}

From the criteria mentioned above to statically define the groups of \humans, \semibots, and \bots, it can be deduced that the number of users within each group is different (Algorithm~\ref{alg:2}).

In particular, we have 709,212 \humans, 208,836 \semibots, and 109,257 \bots. The rest of 270,670 users to complete the full sample of 1,297,975 were not classified due to be suspended or private Twitter profiles.

Therefore, we conclude that our case study is \textbf{unevenly distributed} in three groups of different dimensions.

\subsection{Network analysis}

From a global perspective, we explore three stages of constructing the overall retweet network and their evolution as we increasingly add the three defined groups (Algorithm~\ref{alg:3}). Table~\ref{table:composition} summarizes the graph properties of a) the low-automation legitimate network $\mathcal{G}_{Likely\ Humans}$; b) the intermediate-automation interfered network $\mathcal{G}_{Likely\ Humans \, \cup \, Likely\ SemiBots}$, and c) the high-automation contaminated network $\mathcal{G}_{Likely\ Humans \, \cup \, Likely\ SemiBots \, \cup \, Likely\ Bots}$.

\begin{table}[h!]
    \centering
    \renewcommand{\arraystretch}{1.3}
    \begin{tabular}{|l|c|c|c|}
        \hline
        \multirow{2}{*}{\textbf{Metric}} &
        \multicolumn{3}{c|}{\textbf{Network}} \\
        \cline{2-4} & \textbf{Legitimate} & \textbf{Interfered} & \textbf{Contaminated} \\
       \hline \hline
        Order     & $709,212$ & $918,048$ & $1,027,305$ \\\hline
        Size      & $2,407,485$ & $3,742,115$ & $4,313,219$ \\\hline
        Density   & $4.79 * 10^{-6}$ & $4.44 * 10^{-6}$ & $4.09 * 10^{-6}$\\\hline
        Avg Degree & $3.395$ & $4.076$ & $4.199$ \\\hline
        Avg Strength & $4.460$ & $5.39$ & $5.536$ \\\hline
        Avg Closeness & $0.387$ & $0.377$ & $0.380$ \\\hline
        Avg Betweenness & $1.39 * 10^{-7}$ & $1.32 * 10^{-7}$ & $1.15 * 10^{-7}$\\\hline
        \hline
        \end{tabular}
    \caption{Graph properties of incremental groups}
    \label{table:composition}
\end{table}

In terms of graph order, the network increases along with the size of groups. However, the graph size does not respect a linear continuity, and as we add \semibots\ and \bots, the number of edges (interactions) increases disproportionately. This phenomenon is clearly visible in both average degree and average strength values, which continuously increase in the successive networks. Therefore, the social network becomes compromised with more different connections (edges) and bulks of illegitimate interactions (weights) than expected.

Nevertheless, the proportion of new interactions with respect to new users is not enough to maintain the graph density. Consequently, the retweet network gets sparser (as in normal situations when new users interact in a social network). In terms of centrality, both the normalized average closeness centrality and normalized average betweenness centrality decrease as we add the interfering and contaminant actors, so legitimate users are slightly separated from each other, and more alternative data flows are created.

Generally, from Table~\ref{table:composition}, we can deduce that \semibots\ and \bots\ are \textbf{ecosystem changers} that distort the properties of the network by issuing noise and hindering the natural Twitter dynamics.

\begin{figure}[h!]
    \centering
    \includegraphics[width=\columnwidth]{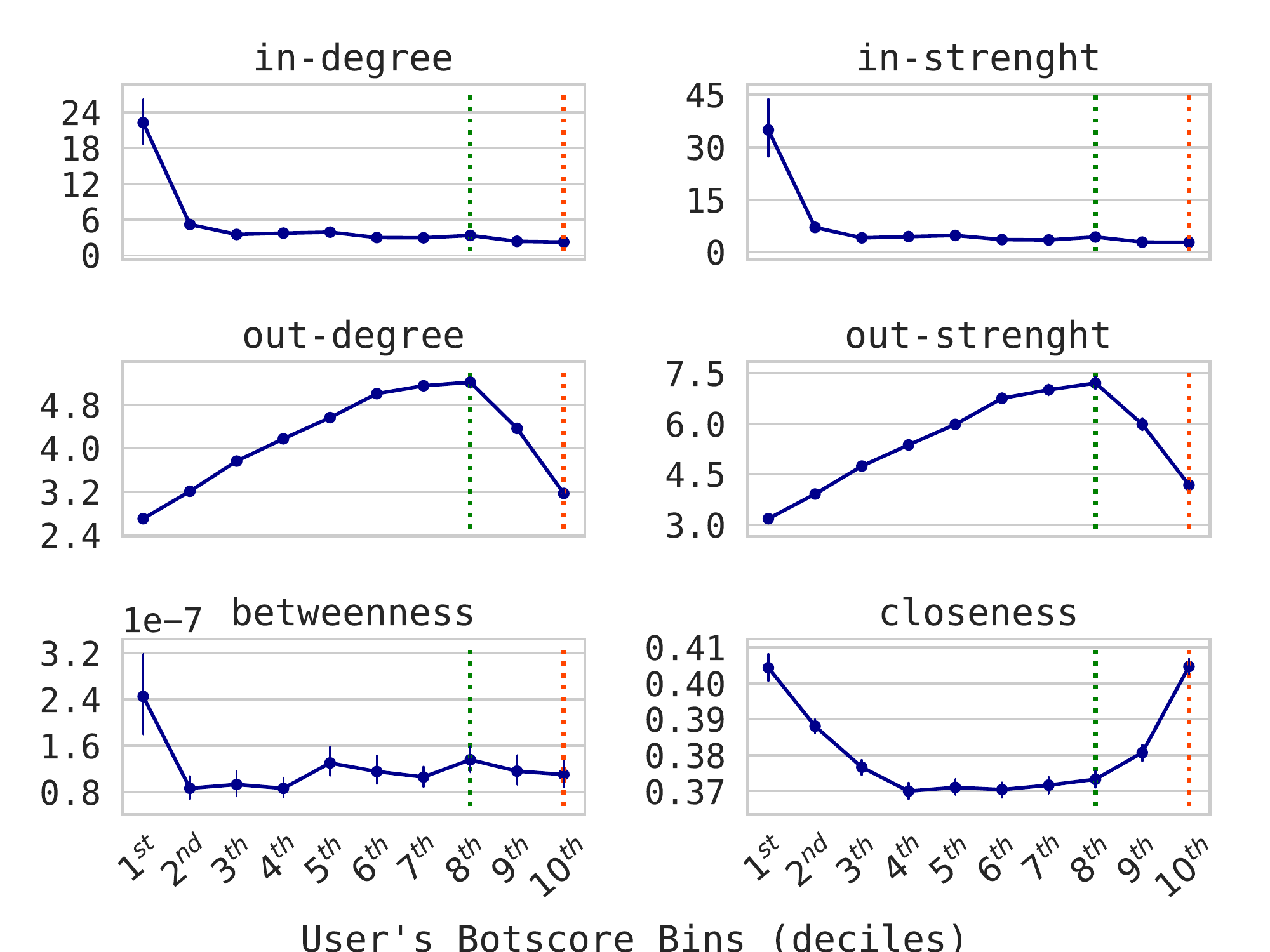}
    \caption{Network composition of groups}
    \label{fig:composition}
\end{figure}

From the node perspective, as opposed to studying the general composition of the social graph, we reveal the in-depth network properties of users (Algorithm~\ref{alg:4}) depending on their botscore. \figureautorefname~\ref{fig:composition} averages the basic metrics per botscore bins (users are grouped to the nearest upper decile) to understand the impact of groups on the network topology (the colored vertical lines divide the charts into the defined groups). 

The in-degree, which reveals the number of different users retweeted, suggests that the 1$^{st}$ decile (\humans) is more \textit{active} retweeting different users (around 20) than the rest. In fact, \semibots\ and \bots\ are the groups that retweet the least number of different users (between 2 and 3). The in-strength values, which consider the weight of edges, follow an identical pattern and reveal the first decile presents the apparent role of \textit{consumers} retweeting over 30 times, unlike the rest, who barely retweet 5 times.

The out-degree of a node, which in our retweets network indicates the number of different users who have retweeted a user, reflects that the peak is achieved by \semibots, who could be considered the most \textit{popular} group. Generally, the higher values are distributed around this peak, attracting between 4 and 5 retweeters. The out-strength follows the same pattern, suggesting that the number of retweets increases with the automation of retweeted accounts from \humans\ to \semibots. The latter could be considered as \textit{producers} whose content is shared between 6 and 8 times. In contrast, \bots\ appear to be much less attractive.

Finally, the betweenness centrality of the 1$^{st}$ decile is the highest, which is great news, meaning that they are \textit{diffusers} in many data paths of the retweet network. Setting aside the first decile, \semibots\ and \bots\ score similar ratings to the rest of \humans. On the other hand, closeness centrality is the most distinguishing feature of the \bots\ compared to the other metrics. Both extremes, normal users (without any automation) and fully automated users, tend to be closer to others and more \textit{visible} and \textit{reachable} through the social network, capable of quickly influencing the rest.

To sum up this perspective of the framework, we conclude that the three defined groups \textbf{behave differently}.

\subsection{Robustness analysis}

\begin{figure}[h!]
    \centering
    \includegraphics[width=\columnwidth]{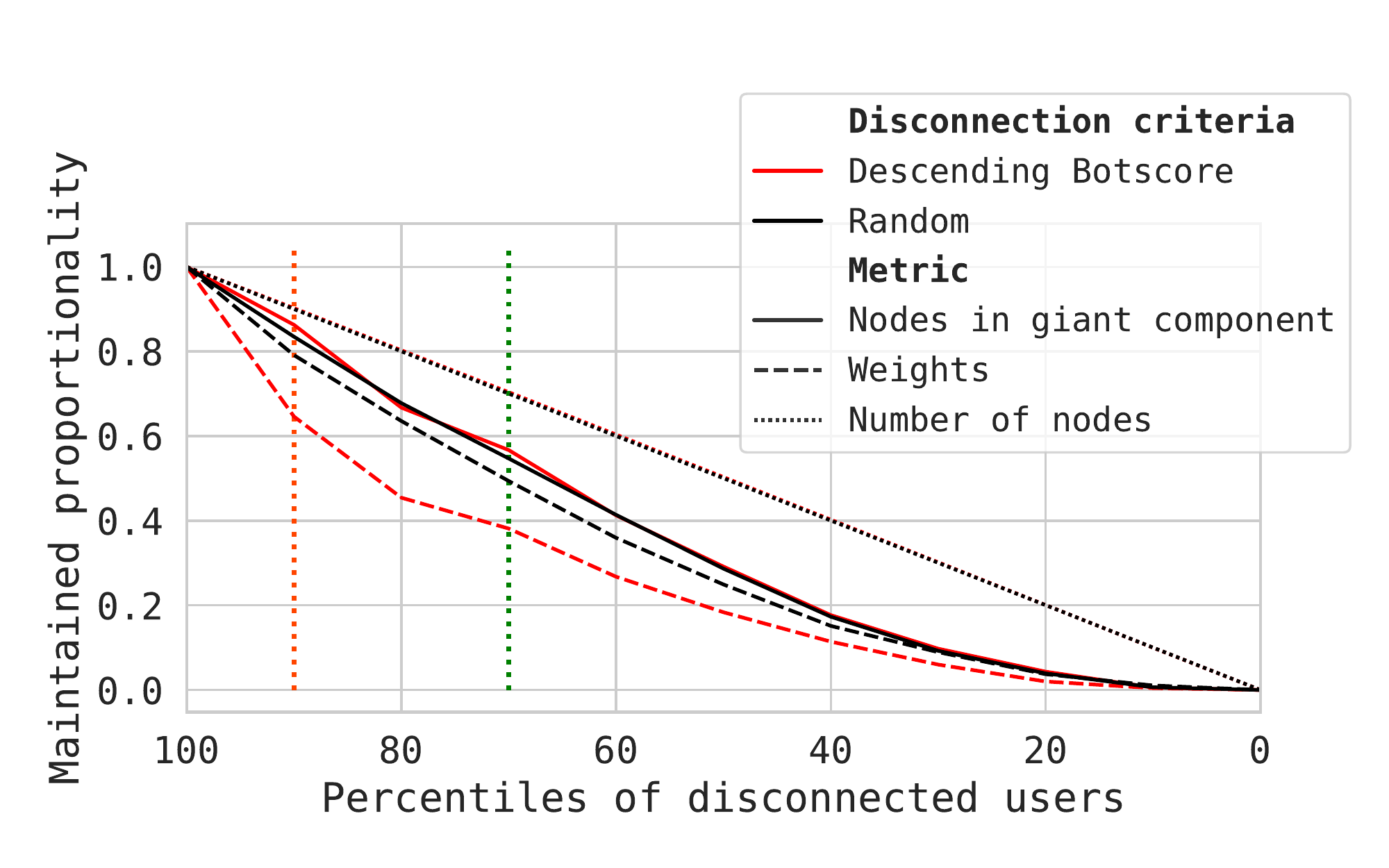}
    \caption{Robustness to the disconnection of users}
    \label{fig:robustness}
\end{figure}

We evaluate the robustness of the network by removing nodes and measuring the associated impact in some network metrics (\algorithmcfname~\ref{alg:5}). \figureautorefname~\ref{fig:robustness} presents the percentage of i) nodes in the graph, ii) nodes in the giant component, and iii) weight of the graph as the percentiles of users are removed. In particular, we repeat the experiment adopting two criteria of disconnecting the percentiles, a) randomly (in black) and b) by botscore in descending order (in red). Therefore, \bots\ are eliminated up to the orange line; \semibots\ are removed up to the green line; and, finally, groups of \textit{Likely-Humans} are disconnected.

Firstly, it is evident that the number of nodes that remains in the graph is inversely proportional to the percentiles removed. 

Secondly, we observe that disconnecting either randomly or by botscore has similar results in the percentage of users in the giant component, decreasing non-linearly. This phenomenon is expected in a sparse retweet network where echo chambers and social microenvironments are connected to the core network by one or a few nodes. As there is no difference with respect to a random disconnection of users, we learn that both \bots\ and \semibots\ are not particularly located at the nexus between the islands and the giant component (which would be a problem for isolated groups). 

Finally, we note that the botscore influences the weight of the network, that is, the total number of retweets. Random disconnection follows a similar pattern than explained previously. However, the weight of the graph falls disproportionately by 40\% just by removing the \bots\ (which represents 10\% of the user sample). This is not the case with the \semibots, which do reduce the total weight proportionally by 20\%. To complete the balance, we see that \humans, 70\% of the users, are responsible for 40\% of the retweets.

Therefore, the network is robust at the level of connectivity to \bots\ and \semibots, who are \textbf{non-destabilizing} users. On the contrary, the scenario is highly sensitive to the number of iterations, where \bots\ are specially \textbf{destabilizing} users.

\subsection{Influence analysis}
 
\begin{figure}[h!]
    \centering
    \includegraphics[width=\columnwidth]{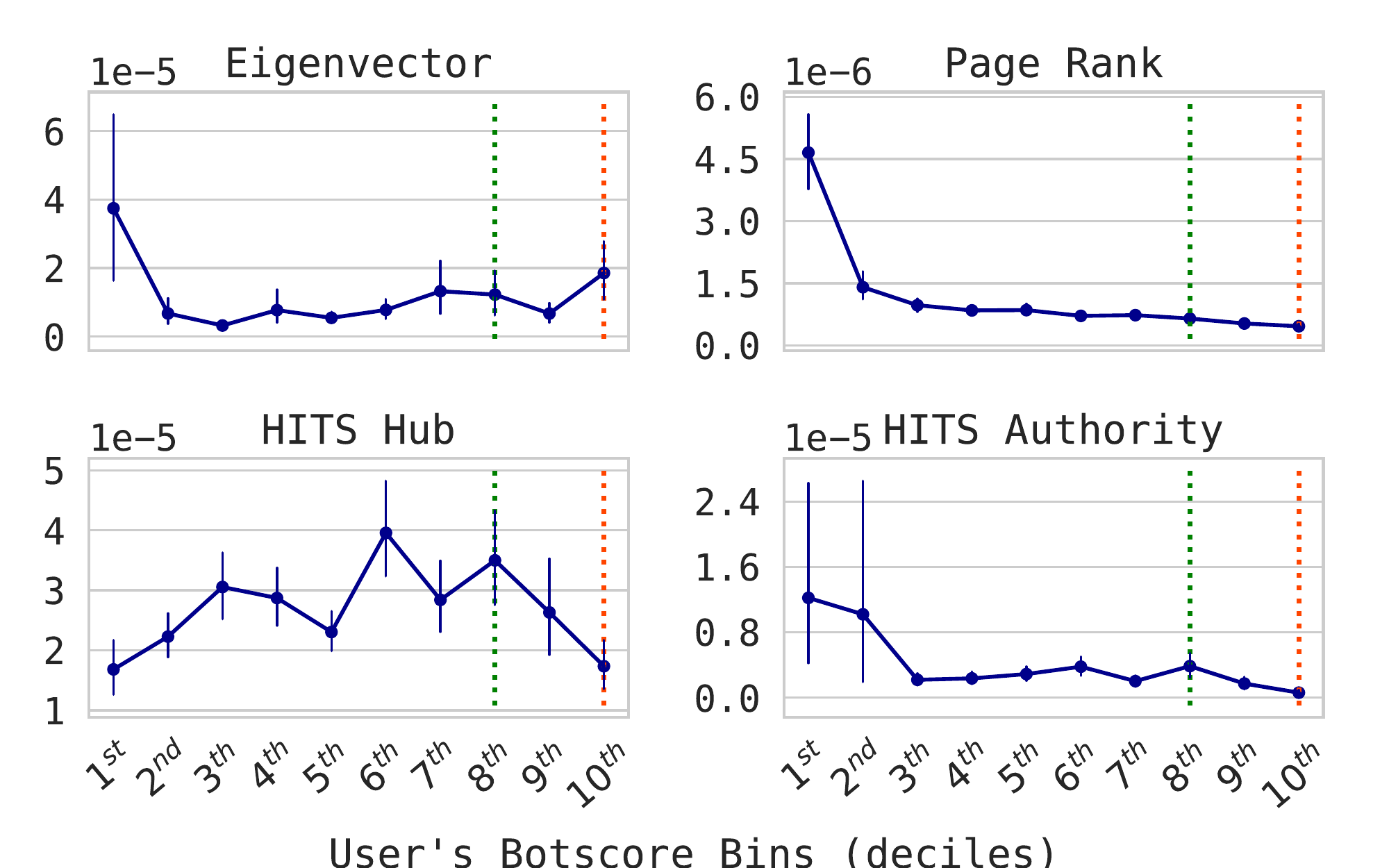}
    \caption{Influence of users through Eigenvector, Page Rank, and HITS algorithms}
    \label{fig:spreading}
\end{figure}

With regard to the spreading capability of the groups, we calculate the well-known centrality metrics of eigenvector, PageRank, and HITS (\algorithmcfname~\ref{alg:6}). \figureautorefname~\ref{fig:spreading} depicts the obtained values grouping the botscores by the nearest upper decile. 

Following the line of the node-based metrics, the eigenvector centrality reveals that in the 1$^{st}$ decile, we find the users who surround themselves with the most active and popular people. Nonetheless, \semibots\ achieve similar results to the rest of \humans, and \bots\ get second place in being well connected.
 
On the other hand, the PageRank centrality decreases with the botscore, suggesting that \humans\ profiles are most likely ``visited" if we follow the propagation of messages through the social graph. This phenomenon is related to the in-degree tendency discussed above.
 
In terms of HITS, the 6$^{th}$ (\humans) and 8$^{th}$ (\semibots) deciles achieve the greatest HITS Hub score, which is related to the out-degree and out-strength (\figureautorefname~\ref{fig:composition}). In contrast, the HITS Authority values are on a downward trend with respect to the botscore, which is linked to the user's in-degree and in-strength. As high HITS authority values are achieved by receiving links from good HITS hubs and vice versa, we deduce that \semibots\ and nearby \humans\ (\textit{hubs} users) are mainly retweeted by the first two deciles of \humans\ (\textit{authority} users).
    
From this experiment, we conclude that the three groups \textbf{influence differently}.

 \subsection{Structure analysis}
 
We measure the presence of the groups in the retweet network from the periphery to the core with the $k$-shell decomposition (\algorithmcfname~\ref{alg:7}). As each $k$-shell has different number of nodes (the layers follow a power-law distribution), we plot the relative presence that each group has in each $k$-shell to facilitate the analysis. 

\begin{figure}[h!]
    \centering
    \includegraphics[width=0.9\columnwidth]{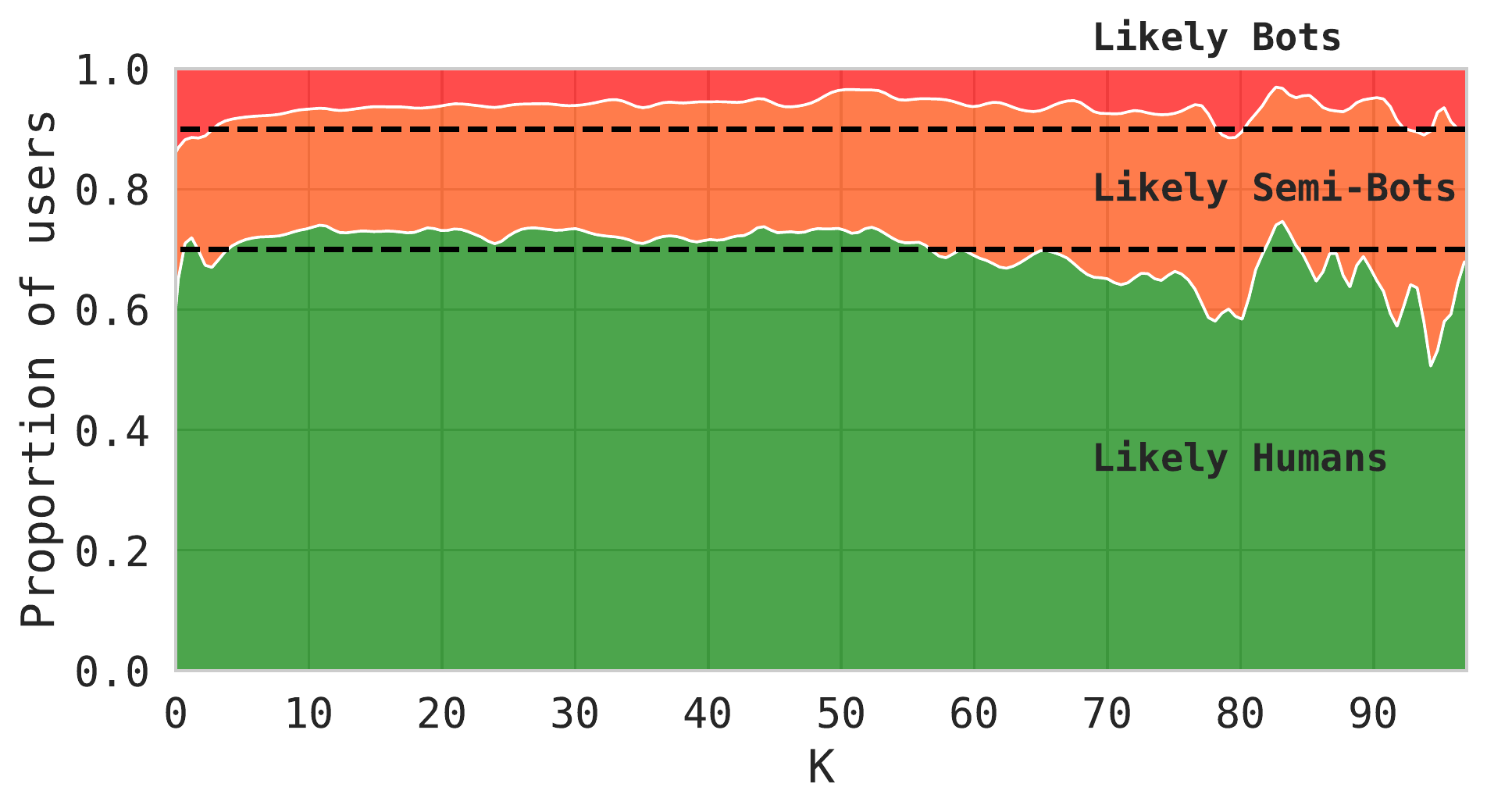}
    \caption{Relative presence of each group in the $k$-shells}
    \label{fig:slicing}
\end{figure}

\figureautorefname~\ref{fig:slicing} shows the kernel density estimation of \humans\ (in green), \semibots\ (in orange), and \bots\ (in red) within the existing $k$-shells, concretely from $k=0$ (minimum) to $k=97$ (maximum).

Intuitively, a balanced distribution of the groups along the network layers would respect the global distribution of users delimited by the dashed lines, \ie 70\% of \humans, 20\% of \semibots, and 10\% of \bots. However, \humans\ are \textbf{highly populating} the surroundings of the network until $k=65$, which leads to a lower number of users from the other two groups. From that $k$-shell, the scenario generally changes in favor of \semibots, who are \textbf{highly populating} the core network. Finally, \bots\ are primarily present at $k=1$, leading the rest of $k$-shells to be \textbf{quite depopulated} of them.

\subsection{Temporal analysis}

\begin{figure}[h!]
    \centering
    \includegraphics[width=0.9\columnwidth]{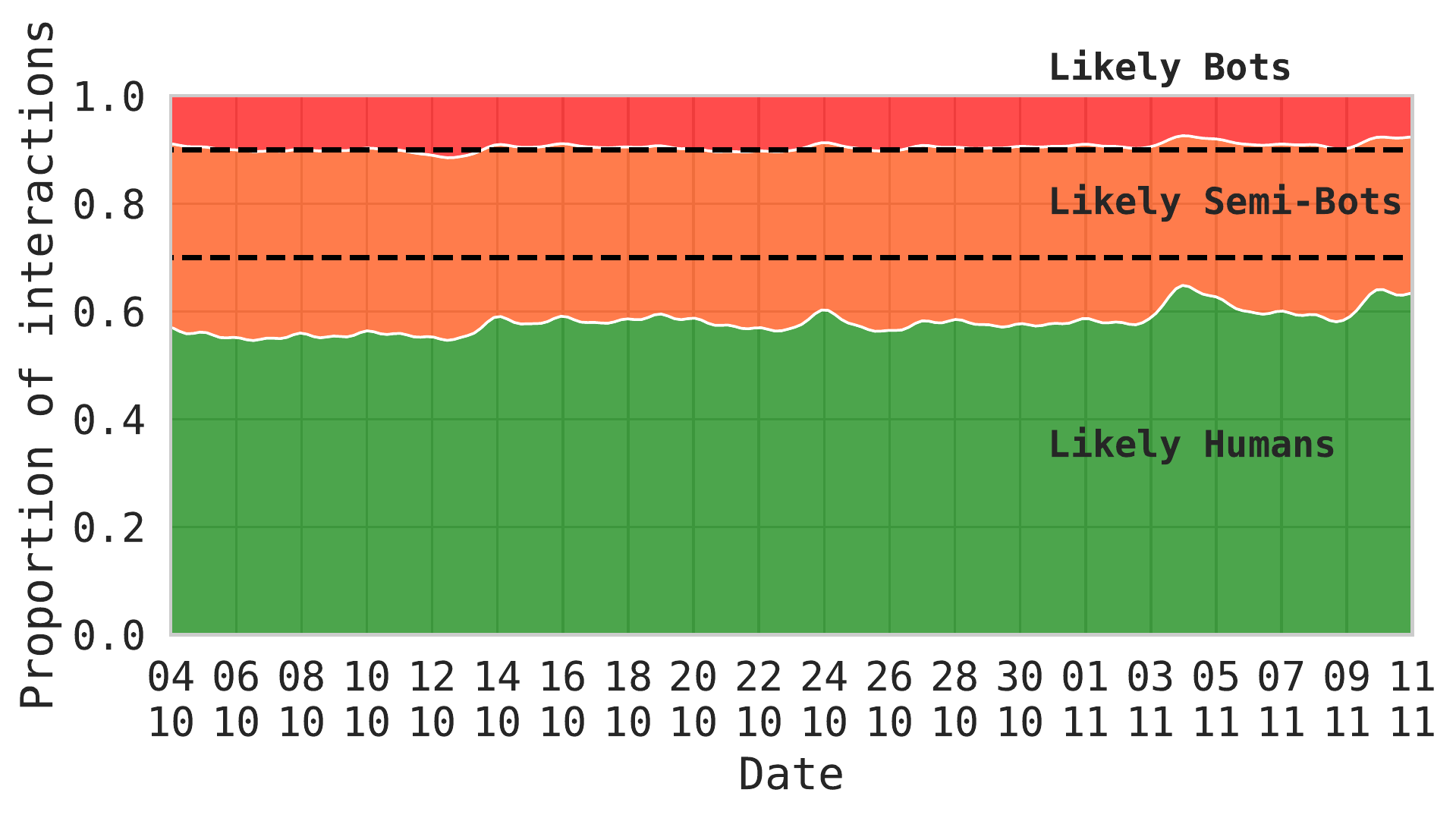}
    \caption{Proportion of daily traffic created by groups}
    \label{fig:timeline}
\end{figure}

Regarding the timeline of the collection window, we are interested in detecting disruptive patterns in the number of interactions (\algorithmcfname~\ref{alg:8}). \figureautorefname~\ref{fig:timeline} shows the proportion of traffic generated by each of the three groups from 04-10-2019 to 11-11-2019.

The trend is practically uniform throughout the days. Considering the dashed lines as a reference of proportional behaviour of groups, we identify that \semibots\ are being \textbf{overstimulated} throughout the time window, having approximately 10\% more activity than expected. As \bots\ do dump their proportional amount of content, the consequence is that \humans\ are \textbf{understimulated} over the four weeks. Nevertheless, the latter tend to have more activity as the election day approaches.

\subsection{Virality analysis}
 
\begin{figure}[h!]
    \centering
    \includegraphics[width=\columnwidth]{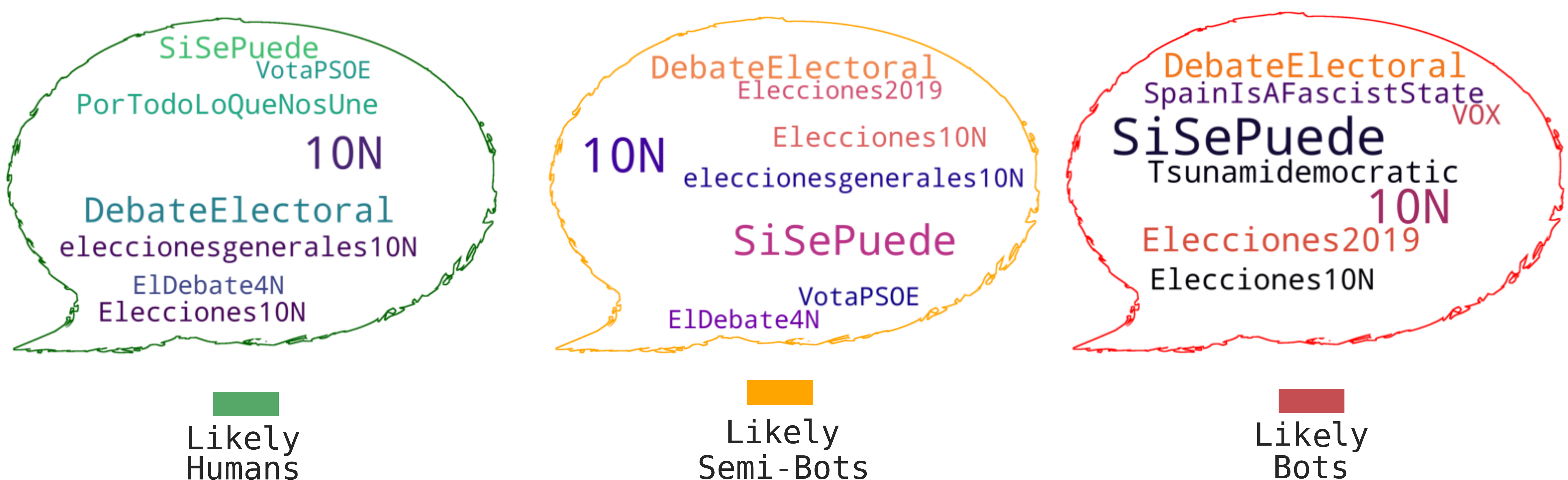}
    \caption{Word clouds of the eight most used hashtags by groups}
    \label{fig:topics}
\end{figure}

Finally, we focus on the created content and its spread over the social graph (\algorithmcfname~\ref{alg:9}). \figureautorefname~\ref{fig:topics} shows the eight most used hashtags within our sampled dataset per group. Generally, we appreciate that the three groups \textbf{discuss similarly} and, unfortunately, have a very similar narrative. They mainly tweeted about the election and the debate on television, although \bots\ also tended to write about the problems in Catalonia.
 
%\begin{figure}[h!]
%    \centering
%    \includegraphics[width=\columnwidth]{img/duration_size.pdf}
%    \caption{Cascades}
%    \label{fig:cascades}
%\end{figure}

% \begin{figure}[h!]
%    \centering
%    \includegraphics[width=\columnwidth]{img/cascades-average.pdf}
%    \caption{Cascades}
%    \label{fig:cascades}
%\end{figure}

\begin{figure}[h!]
    \centering
    \includegraphics[scale=0.49]{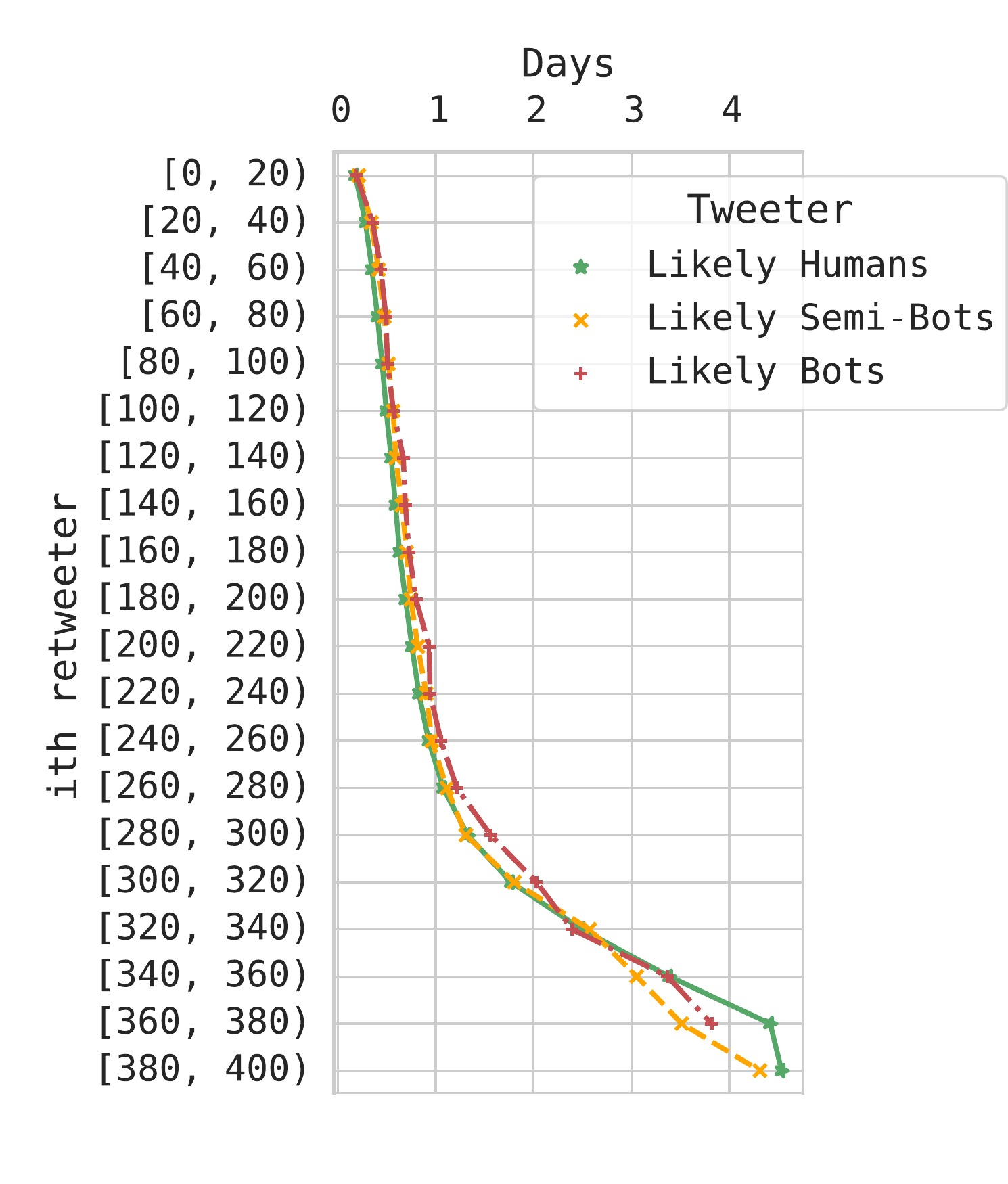}
    \caption{Information cascades of retweets over time}
    \label{fig:cascades}
\end{figure}

According to the spread of the tweets, \figureautorefname~\ref{fig:cascades} shows the average number of days it takes for the $i$-th retweeter to share a message, differentiating by the tweeter group. The diffusion cascades formed have a logarithmic evolution regardless of the creator of the tweet, eventually getting between 240 and 260 shares over the first day of the tweet. The retweet frequency decreases significantly from the first day on-wards, reaching approximately 60 retweets on the second day, 40 retweets on the third day, and 20 on the fourth and fifth day, respectively. To highlight a slight difference, it is curious that \semibots\ take less time to achieve between 340 and 400 retweets than the other two groups.

Therefore, the three groups are \textbf{non-influencers} as they do not provoke extensive cascades in short periods and are \textbf{equally viral} since they undergo similar diffusion patterns.

%% file: discussion.tex
\section{Overview of social bots interference}
\label{discussion}

\begin{table*}[h!]
    \centering
    \renewcommand{\arraystretch}{1.2}
    \begin{tabular}{|l|c|c|c|}
        \hline
        \multicolumn{1}{|l}{\textbf{Analysis perspective}} & \multicolumn{3}{|c|}{\textbf{Output}} \\
       \hline \hline
        Groups definition          & \textit{\humans\ (70\%)} & \textit{\semibots\ (20\%)} & \textit{\bots\ (10\%)} \\\hline
        Statistical    &  \multicolumn{3}{c|}{\textit{Unevenly distributed}}     \\\hline
        Network (global)  & \textit{Base} &  \multicolumn{2}{c|}{\textit{Ecosystem changers}}     \\\hline
        Network (nodes)    & \multicolumn{3}{c|}{\textit{Behave differently}}     \\\hline
        Robustness      & \textit{Base} & \textit{Non-destabilizing} & \textit{Destabilizing}    \\\hline
        Influence       &  \multicolumn{3}{c|}{\textit{Influence differently}}     \\\hline
        Structure         & \makecell{\textit{Populating}\\\textit{surroundings}} & \makecell{\textit{Populating}\\\textit{core}} & \makecell{\textit{Populating}\\\textit{K=1}}    \\\hline
        Temporal        & \textit{Understimulated} & \textit{Overstimulated} &  \textit{Normal}   \\\hline
        Virality        & \multicolumn{3}{c|}{\textit{Discuss similarly, non-influencers and equally viral}}     \\\hline
        \hline
        \end{tabular}
    \caption{Framework output for the analysis of social bots within the 2019 Spanish election}
    \label{table:results}
\end{table*}

The experiments described in Section~\ref{experiments} reveal both interesting quantitative results as well as qualitative characterizations. We present the overview of the framework output in \tableautorefname~\ref{table:results}.

In the definition of groups, we start from the hypothesis that we have three types of users: \humans\ (natural users), \semibots\ (profiles mixing manual and automated behaviour), and \bots\ (programmed accounts). They are unevenly distributed, representing 70\%, 20\%, and 10\% of the sample, respectively. 

By studying the global composition of the network, we discover that the social graph properties change as we add the other two groups to the base network of \humans. Therefore, if \bots\ and \humans\ are ecosystem changers, they are somehow different. 

The inspection of the network from the node perspective confirms that groups behave differently. The first decile of \humans\ (potentially verified or \textit{influencers}) are the most retweeting users and are in the middle of many data paths. \semibots\ are successful in being retweeted, which could be those troll, hater, or parody accounts generating witty content to laugh at the political parties. On the other hand, we discover that \bots\ are strategically located to be easily accessible to the rest of the users. Fortunately, their highly automated tweets are not highly shared, which could indicate that they are too artificial. Perhaps, \bots\ are those informative or joking accounts without any specific purpose, which by their nature usually appear in most of the timelines, but do not manage to become viral.

In terms of connectivity, the social network seems to be reasonably resistant to the disconnection of malicious nodes. Therefore, the network is stable, and it is not easy for groups to manipulate it by creating malicious social bubbles. This fact was also highlighted in the global composition exploration, where the averaged betweenness and closeness centralities were reduced in the interfered and contaminated network. However, the robustness experiments prove that \bots\ are involved in more tweets than expected, destabilizing natural Twitter dynamics.

Assessing the diffusion capacity and their neighbourhood, we learn that each group influences differently, and the first decile (\humans) stands out again. The eigenvector centrality suggests that \semibots\ tend to interact with more engaging users than \humans. Moreover, \semibots\ not only attract lots of interactions as indicated by the out-strength, but HITS confirms as well that they do it from those who retweet a lot.

The inspection of the snapshot as an onion reaffirms that \semibots\ are highly communicated with each other and with other super-spreaders. Their popularity and content production (out-degree and out-strength), their position in the diffusion (betweenness), and their connection with other good nodes (eigenvector) cause them to highly populate the core network.  On the other hand, the timeline view confirms that semi-automated accounts are overstimulated during the monitoring window, provoking more interactions than expected. Nevertheless, we can extract no anomalous patterns or suspicious peaks on any of the days for any group.

Last but not least, the three groups discussed basically the same topics, mimicking each other. Therefore, both \bots\ and \semibots\ could aim to invade trending hashtags. Although the topology of the network favours \semibots\ by being connected with powerful users, and they generate proportionally more content than the rest, they do not become viral or create trends. Optimistically, it is positive that they do not monopolize the diffusion of tweets. Pessimistically, they have a better context than the rest and the same capacity as \humans\ in spreading. Hopefully, this is a everyday occurrence, as these actors could exploit the social network algorithms (and their weaknesses) to their advantage. At the same time, legitimate users make the expected use of the social network.

%% file: conclusions.tex
\section{Conclusions and future work}
\label{conclusions}
Existing work to date on the study of bots has focused on modeling detection tools and using them to inspect specific scenarios. In the absence of clearly defined methodologies to characterize these accounts and calculate their relative impact with respect to legitimate groups, we have proposed BOTTER, a framework for analyzing social bots in Twitter.

Based on the group of users to be compared (usually humans and bots), our framework defines seven analysis perspectives (statistical, network, robustness, influence, structure, temporal, content, and virality). The latter return metrics for group comparison, highlighting their differences and measuring the interferences between them. Our framework relies on data computation through widely adopted algorithms, guaranteeing objective and non-biased results. 

On the other hand, we have successfully applied and validated the framework with a Twitter dataset collected in the weeks immediately before the 2019 Spanish election. From the definition of three groups (\humans, \semibots\ and \bots), the experimental results indicate that the analyzed snapshot was difficult to manipulate in terms of its network structure. However, semi and fully automated agents certainly changed the natural dynamics of the network. We found that semi-automated accounts may have grabbed the attention of users and succeeded in being connected to the most active and popular nodes of the network. This fact would potentially worry if those accounts used their advantageous position in the network to break its integrity. Fortunately, while they were capable of generating many interactions, we did not detect them going viral or overly influential. In contrast, the fully automated bots stood out in being close to users and easily reachable within the network, even attracting interactions. However, they were located on the periphery, and their content did not spread noticeably. Therefore, the social bots would be altering the social network but not succeeding in their influencing tasks. Nevertheless, it is worth noting that semi-automated bots seem to have a much more significant impact than fully automated ones.

Therefore, the validated solution is mature to guide future research on this issue, enabling the direct comparison of results from studies that eventually adopt it. The perspectives are designed in a decoupled way to adopt the desired ones (if not all of them are of interest) and are applicable to compare any Twitter group of users (not only bots). 

Nevertheless, our proposal deals more with the horizontal definition of the framework rather than the depth. The analysis perspectives include the de facto standard metrics and algorithms widely used in the literature, but they can be further developed with more sophisticated metrics or algorithms. In particular, it would be interesting to include i) group-to-group interaction relationships and aggregated account profile features (within static analysis), and ii) sentiment and message quality/veracity analysis (within content analysis). Another line of future work would be to introduce an interpretative guide to the results of each perspective for facilitating qualitative discussion. Finally, it would be interesting to extrapolate the framework to other social networks.